\documentclass[12pt, a4paper]{article}
\usepackage[english]{babel}
\usepackage{verbatim}
\usepackage{bm}

\usepackage{amsmath}
\usepackage{amssymb}
\usepackage{amsfonts}
\usepackage{amsbsy}

\usepackage{graphicx}
\usepackage[utf8]{inputenc}

\usepackage{newlfont}
\usepackage{float}
\usepackage{siunitx}
\usepackage[paperwidth=192mm,
			paperheight=262mm,
			vmargin={19mm,19mm},
			hmargin={13.7mm,13.7mm},
			headsep=12pt,
			footskip=12pt]{geometry}

\usepackage{hyperref}

\DeclareMathOperator{\grad}{\nabla}
\DeclareMathOperator{\dive}{\nabla\cdot}
\DeclareMathOperator{\dives}{\nabla_s\cdot}

\usepackage{subcaption}

\begin{document}

\title{On the evolution equations of interfacial variables in two-phase flows}

\author{Giuseppe Orlando$^{(1)}$ \\ 
		Paolo Francesco Barbante$^{(1)}$, Luca Bonaventura$^{(1)}$}

\date{}

\maketitle

\begin{center}
{\small
$^{(1)}$  
MOX, Dipartimento di Matematica, Politecnico di Milano \\
Piazza Leonardo da Vinci 32, 20133 Milano, Italy\\
{\tt giuseppe.orlando@polimi.it, paolo.barbante@polimi.it, luca.bonaventura@polimi.it}
}
\end{center}


\pagebreak

\abstract{Many physical situations are characterized by interfaces with a non trivial shape so that relevant geometric features, such as interfacial area, curvature or unit normal vector, can be used as main indicators of the topology of the interface. We analyze the evolution equations for a set of geometrical quantities that characterize the interface in two-phase flows. Several analytical relations for the interfacial area density are reviewed and presented, clarifying the physical significance of the different quantities involved and specifying the hypotheses under which each transport equation is valid. Moreover, evolution equations for the unit normal vector and for the curvature are analyzed. The impact of different formulations is then assessed in numerical simulations of rising bubble benchmarks.}

\pagebreak

\section{Introduction}
\label{sec:intro} \indent

Two-phase flows play an important role in several natural processes and engineering systems. A moving interface delimits the bulk regions of the single phases. According to the interface geometrical configuration, two-phase flows can be classified as either separated or disperse. A separated flow is characterized by large regions of both phases. On the contrary, a disperse flow is constituted of particles, such as bubbles or droplets, dispersed in a carrier fluid, which is called the continuous phase. In both separated and disperse flows, the exchanges between the two phases occur at the interface and phase exchange terms are proportional to the interface area \cite{drew:1999, seroguillaume:2005}. Hence, the computation of this quantity is a prerequisite to obtain reliable values of exchange terms, especially in non-equilibrium conditions or when chemical reactions occur. The use of suitable evolution equations to predict the interfacial area concentration has a long tradition in the literature, see e.g. \cite{drew:1990, drew:1999, hibiki:2002, ishii:1975}, and represents, in the case of disperse flows, a popular alternative to Population Balance Equation (PBE) models, like that proposed in \cite{williams:1958}, which applies the method of moments to derive several transport equations for the moments of the considered Probability Density Function (PDF) \cite{mcgraw:1997}. The use of a relation for the evolution of the interfacial area leads instead to a single transport equation, thus providing a significant advantage in terms of computational efficiency with respect to the alternative PBE approach. Moreover, since the phase exchange terms depend on the geometry of the interface, considering higher order statistics such as the mean curvature or the outward unit normal vector could significantly improve the description of such terms.
  
The analysis of evolution equations of geometrical quantities that characterize the interface in two-phase flows has been subject of several contributions, see among many others \cite{drew:1990, lhuillier:2003, lhuillier:2004, morel:2007}. Most of these investigations were focused on the analysis of the evolution equation for the interfacial area density, which will be reviewed and extended in Section \ref{sec:interface_area}. The evolution equations for the unit normal vector and the mean curvature have been analyzed in \cite{drew:1999}, considering only the normal component of the interfacial velocity. In this work, following the discussion in \cite{orlando:2023a}, we will extend these relations considering the complete interfacial velocity as advecting field, providing thus more generality in the analysis of the dynamics of the interface. The two phases exchange mass, momentum, and energy across the interface, and a suitable modelling for such terms is therefore needed. In future work, we plan to couple the equations developed in Section \ref{sec:interface_area} and \ref{sec:normal_curvature} with the multifluid Bear-Nunziato type equations for two-phase flows. The results presented in Section \ref{sec:numerical} show that the equations developed in this article can be used to predict the interfacial area.

The paper is structured as follows. In Section \ref{sec:local_balance_laws}, we review the classical local balance laws which model two-phase flows and we show that the interfacial source terms are proportional to the interfacial area density. In Section \ref{sec:interface_area}, we obtain local instantaneous transport equations for the interfacial area density. In Section \ref{sec:normal_curvature}, we analyze the evolution of the unit normal and of the mean curvature. Some numerical experiments to validate the different relations are proposed in Section \ref{sec:numerical}. Some conclusions and perspectives for future work are finally presented in Section \ref{sec:conclu}.

\section{Local balance equations}
\label{sec:local_balance_laws} \indent

Let \(\Omega \subset \mathbb{R}^{d}, 2 \le d \le 3\) be a connected open bounded set with a sufficiently smooth boundary \(\partial\Omega\). The canonical form of a balance equation can be written as \cite{drew:1999}
\begin{equation}\label{eq:generic_balance}
\frac{\partial \rho \Psi}{\partial t} + \dive \left(\rho \Psi \mathbf{u}\right) = \dive \mathbf{J} + \rho f \qquad \mbox{in } \Omega.
\end{equation}
Herein \(\Psi\) is the conserved quantity (either a scalar or a tensorial one), \(\rho\) is the density, \(\mathbf{u}\) is the velocity, \(\mathbf{J}\) is the flux (molecular or diffusion) of the variable \(\Psi\) and \(f\) is the source density. For \(\Psi = 1, \mathbf{J} = 0\) and \(f = 0\), we obtain the continuity equation
\begin{equation}\label{eq:conservation_mass}
\frac{\partial \rho}{\partial t} + \dive\left(\rho \mathbf{u}\right) = 0 \qquad \mbox{in } \Omega. 
\end{equation} 
For \(\Psi = \mathbf{u}, \mathbf{J} = \mathbf{T}\) and \(f = \mathbf{b}\), we get the balance of momentum
\begin{equation}\label{eq:balance_momentum}
\frac{\partial \rho \mathbf{u}}{\partial t} + \dive\left(\rho \mathbf{u} \otimes \mathbf{u}\right) = \dive \mathbf{T} + \rho \mathbf{b} \qquad \mbox{in } \Omega.
\end{equation} 
Here \(\mathbf{T}\) is the stress tensor and \(\mathbf{b}\) is the body force. Eventually, for \(\Psi = E, \mathbf{J} = \mathbf{T}\mathbf{u} - \mathbf{q}\) and \(f = \mathbf{b} \cdot \mathbf{u} + r_{heat}\), we obtain the balance of energy
\begin{equation}\label{eq:balance_energy}
\frac{\partial \rho E}{\partial t} + \dive\left(\rho E \mathbf{u}\right) = \dive \left(\mathbf{T}\mathbf{u} - \mathbf{q}\right) + \rho \left(\mathbf{b} \cdot \mathbf{u} + r_{heat}\right) \qquad \mbox{in } \Omega,
\end{equation} 
where \(E\) is total energy per unit of mass, \(\mathbf{q}\) is the heat flux and \(r_{heat}\) is the heating source per unit mass. \\
From now on, we assume that the domain \(\Omega\) consists of two subdomains \(\Omega_{1}(t)\) and \(\Omega_{2}(t)\), separated by an interface \(\Gamma(t)\). Hence, we consider a single discontinuity across a smooth surface separating two parts occupied by the fluid where the fields are smooth. We aim to analyze the aforementioned physical laws directly incorporating the interface conditions inside the balance equations. To do so, we define the characteristic function \(X_{k}\) of phase \(k\) as 
\begin{equation}\label{eq:characteristic_function_def}
X_{k}(\mathbf{x},t) = 
\begin{cases} 
	1, &\mbox{if \(\mathbf{x}\) is in phase \(k\) at time \(t\)} \\ 
	0 &\mbox{otherwise} 
\end{cases}
\end{equation}
and we take the product of the equation \eqref{eq:generic_balance} with the characteristic function \eqref{eq:characteristic_function_def} so as to obtain
\begin{equation}\label{eq:balance_times_Xk}
X_{k} \frac{\partial \rho\Psi}{\partial t} + X_{k} \dive\left(\rho\Psi\mathbf{u}\right) = X_{k} \dive\mathbf{J} + X_{k}\rho f.
\end{equation}
Products including the characteristic functions are discontinuous, so that the derivatives must be treated in a distributional sense \cite{drew:1999}. If we take the characteristic function inside the derivatives we get
\begin{eqnarray}\label{eq:balance_times_Xk_cons}
&&\frac{\partial(X_{k}\rho\Psi)}{\partial t} + \dive\left(X_{k}\rho\Psi\mathbf{u}\right) - \dive\left(X_{k}\mathbf{J}\right) - X_{k}\rho f = \nonumber \\
&&\rho\Psi\left(\frac{\partial X_{k}}{\partial t} + \mathbf{u} \cdot \grad X_{k}\right) - \mathbf{J} \cdot \grad X_{k}.
\end{eqnarray}
Notice that, with a slight abuse of notation, we employ the same symbol for both distributional and classical derivatives and the proper operator to be considered will follow from the context. Let us briefly analyze the right-hand side of \eqref{eq:balance_times_Xk_cons}, which can be rewritten as
\begin{eqnarray}\label{eq:balance_rhs}
&&\rho\Psi\left(\frac{\partial X_{k}}{\partial t} + \mathbf{u} \cdot \grad X_{k}\right) - \mathbf{J} \cdot \grad X_{k} = \nonumber \\
&&\rho\Psi\left(\frac{\partial X_{k}}{\partial t} + \mathbf{u} \cdot \grad X_{k} + \mathbf{v}_{I} \cdot \nabla X_{k} - \mathbf{v}_{I} \cdot \grad X_{k}\right) - \mathbf{J} \cdot \nabla X_{k} = \nonumber \\
&& \left[\rho\Psi\left(\mathbf{u} - \mathbf{v}_{I}\right) - \mathbf{J}\right] \cdot \grad X_{k}.
\end{eqnarray}
The last equality stems from the so-called topological equation for phase \(k\), which rules the evolution of the characteristic function \cite{drew:1999, junqua-moullet:2003, morel:2007}:
\begin{equation}\label{eq:topological_equation}
\frac{\partial X_{k}}{\partial t} + \mathbf{v}_{I} \cdot \grad X_{k} = 0,
\end{equation}
where \(\mathbf{v}_{I}\) is the interfacial velocity, whose definition will be specified in Section \ref{sec:interface_area}. We substitute \eqref{eq:balance_rhs} into \eqref{eq:balance_times_Xk_cons} so as to obtain
\begin{equation}\label{eq:balance_equation_distributional}
\frac{\partial(X_{k}\rho\Psi)}{\partial t} + \dive\left(X_{k}\rho\Psi\mathbf{u}\right) - \dive\left(X_{k}\mathbf{J}\right) - X_{k}\rho f =  \left[\rho\Psi\left(\mathbf{u} - \mathbf{v}_{I}\right) - \mathbf{J}\right] \cdot \nabla X_k.
\end{equation}
The right-hand side of \eqref{eq:balance_equation_distributional} is the total flux of the property \(\Psi\) of phase \(k\) across the interface \(\Gamma\). As outlined in \cite{junqua-moullet:2003, morel:2007}, \(\nabla X_{k}\) is the product between \(\delta(\Gamma)\), the Dirac delta distribution with support on the interface, and the outward unit normal from phase \(k\) \(\mathbf{n}_{k}\), namely
\begin{equation}\label{eq:grad_Xk}
\nabla X_{k} = -\mathbf{n}_{k}\delta(\Gamma).
\end{equation}
The distribution \(\delta(\Gamma)\) is defined by the following relation:
\begin{equation}
<\delta(\Gamma), \phi> = \int_{\Gamma}\phi d\Sigma \qquad \forall \phi \in C^{\infty}_{0}(\Omega)
\end{equation}
and represents an interfacial area density. Indeed, with a slight abuse of notation, the following relation holds:
\begin{equation}
\int_{\Omega}\delta(\Gamma) d\Omega = A,
\end{equation}
where \(A\) is the interface area. Hence, we remark that the interfacial flux terms on the right-hand side of \eqref{eq:balance_equation_distributional} are proportional to the interfacial area density, whose computation is therefore fundamental for an accurate estimate of these exchange terms.

\section{Local instantaneous transport equations for the interfacial area density}
\label{sec:interface_area} \indent

In this Section, we analyze the evolution equations for the interfacial area density. We assume that the interface \(\Gamma(t)\) between two phases is a \(d-1-\)dimensional manifold in a \(d\)-dimensional Euclidean space with \(d = 2,3\). Notice that we assume that the surface \(\Gamma(t)\) is closed and without contact lines. Two representations of an interface in space can be considered \cite{aris:2012, morel:2007}. In the first description, the surface is seen as the zero-level of a suitable function \(F(\mathbf{x},t)\), with \(\mathbf{x} \in \Omega\) denoting the spatial coordinates and \(t \in \left(0, T_{f}\right]\) denoting the temporal coordinate. Here, \(T_{f}\) is the final time. The second representation is given by
\begin{equation}
\mathbf{x} = \mathbf{x}(\boldsymbol{\alpha},t),  
\end{equation}
where \(\boldsymbol{\alpha}\) are the surface coordinates. The velocity of a point on the surface with coordinates \(\boldsymbol{\alpha}\) is defined as
\begin{equation}
\mathbf{v}_{I} = \frac{\partial\mathbf{x}}{\partial t}.
\end{equation}
In the following, for the sake of simplicity, we omit the explicit dependence on space and time for the different geometric variables. Since \(F\) is identically zero for all the points located on the interface, its Lagrangian derivative at velocity \(\mathbf{v}_{I}\) is null, which entails the following kinematic equation:
\begin{equation}\label{eq:F_evolution}
\frac{\partial F}{\partial t} + \mathbf{v}_{I} \cdot \nabla F = 0.
\end{equation}
Moreover, the unit vector normal to the surface is given by:
\begin{equation}\label{eq:normal_F}
\mathbf{n} = \pm\frac{\nabla F}{\left|\nabla F\right|}.
\end{equation}
From \eqref{eq:F_evolution} and \eqref{eq:normal_F}, it follows immediately that two different interfacial velocity fields with the same normal component yield the same Lagrangian derivative for \(F\). Indeed, if we substitute \eqref{eq:normal_F} into \eqref{eq:F_evolution}, we obtain
\begin{equation}\label{eq:F_evolution_normal_velocity}
\frac{\partial F}{\partial t} \pm \left(\mathbf{v}_{I} \cdot \mathbf{n}\right)\left|\nabla F\right| = 0.
\end{equation}
As already reported in Section \ref{sec:local_balance_laws}, the characteristic function \(X_{k}\) \eqref{eq:characteristic_function_def} follows the same dynamics \eqref{eq:topological_equation}.

We derive now the local instantaneous equations for the Dirac delta distribution \(\delta(\Gamma)\) with support on the interface. For this purpose, we take the gradient of the evolution equation of the characteristic function \eqref{eq:topological_equation} and we obtain the following relation: 
\begin{equation}\label{eq:grad_Xk_evolution}
\frac{\partial \nabla X_{k}}{\partial t} + \nabla\left(\mathbf{v}_{I} \cdot \nabla X_{k}\right) = \mathbf{0}. 
\end{equation}
There are functions, like the outward unit normal vector or the interfacial velocity, whose definitions are properly meaningful only for the points on the surface \(\Gamma\), as explained in \cite{nadim:1996}. Therefore, for these quantities, we have to employ derivatives that are defined intrinsically on the surface. The relations for time and space derivatives of this kind for a generic scalar function \(f\) have been introduced in \cite{estrada:1980, estrada:1985}:
\begin{eqnarray}
\frac{\partial_{s} f}{\partial t} &=& \frac{\partial \tilde{f}}{\partial t} + \left(\mathbf{v}_{I} \cdot \mathbf{n}\right)\left(\nabla\tilde{f} \cdot \mathbf{n}\right) \label{eq:time_surface_derivative_product} \\
\frac{\partial_{s} f}{\partial x_{i}} &=& \frac{\partial\tilde{f}}{\partial x_{i}} - n_{i}\left(\nabla\tilde{f} \cdot \mathbf{n}\right), \label{eq:space_surface_derivative_product}
\end{eqnarray}
where \(\tilde{f}\) is any smooth extension of \(f\) outside \(\Gamma\). In particular, for a first-order tensor \(\mathbf{f}\) we have
\begin{equation}
\dives\mathbf{f} = \left(\mathbf{I}-\mathbf{n}\otimes\mathbf{n}\right):\nabla\tilde{\mathbf{f}} = \dive\tilde{\mathbf{f}} - \left[\left(\nabla\tilde{\mathbf{f}}\right)\mathbf{n}\right]\cdot\mathbf{n}.
\end{equation}
Here, \(\mathbf{I}\) is the \(d \times d\) identity tensor and we define the gradient of a first-order tensor \(\tilde{\mathbf{f}}\) as the second-order tensor whose components are
\begin{equation}
\left[\nabla\tilde{\mathbf{f}}\right]_{ij} = \frac{\partial \tilde{f}_{i}}{\partial x_{j}}.
\end{equation}
Notice that, as explained in \cite{estrada:1985}, given \(f\) defined only on \(\Gamma\), \(f\delta(\Gamma)\) denotes the distribution \(\tilde{f}\delta(\Gamma)\). It is useful that \(\tilde{f}\) satisfies the condition
\begin{equation}\label{eq:null_extension}
\nabla\tilde{f} \cdot \mathbf{n}\bigg\rvert_\Gamma = 0.  
\end{equation}
Analogous considerations hold in the case \(\mathbf{f}\delta(\Gamma)\), for which one can impose
\begin{equation}\label{eq:null_extension_vector}
\left[\left(\nabla\tilde{\mathbf{f}}\right)\mathbf{n}\right] \bigg\rvert_\Gamma = \mathbf{0}.  
\end{equation}
The smooth extensions \(\tilde{f}\) and \(\tilde{\mathbf{f}}\) do not satisfy in general \eqref{eq:null_extension} and \eqref{eq:null_extension_vector}, respectively. However, relations \eqref{eq:null_extension} and \eqref{eq:null_extension_vector} make the value of the surface derivatives independent of the value of \(\tilde{f}\) or \(\tilde{\mathbf{f}}\) so that we can consider \(f\) and \(\tilde{f}\) or \(\mathbf{f}\) and \(\tilde{\mathbf{f}}\) without distinction. On the other hand, if \(f\) or \(\mathbf{f}\) are already defined and regular in the whole space-time domain \(\Omega \times \left(0, T_{f}\right]\), one can define \(f\delta(\Sigma)\) or \(\mathbf{f}\delta(\Sigma)\), but this distribution will also depend on the value of \(\nabla f \cdot \mathbf{n}\bigg\rvert_\Gamma\) or of \(\left[\left(\nabla\mathbf{f}\right)\mathbf{n}\right]\bigg\rvert_\Gamma\). The following relation holds:
\begin{equation}\label{eq:distro_product_rule}
\frac{\partial\left(\tilde{f}\delta(\Gamma)\right)}{\partial\square} = \frac{\partial\tilde{f}}{\partial\square}\delta(\Gamma) + \frac{\partial\delta(\Gamma)}{\partial\square}\tilde{f},
\end{equation}
which reduces to \cite{estrada:1985}
\begin{equation}\label{eq:distro_product_rule_extension}
\frac{\partial\left(\tilde{f}\delta(\Gamma)\right)}{\partial\square} = \frac{\partial_{s}f}{\partial\square}\delta(\Gamma) + \frac{\partial\delta(\Gamma)}{\partial\square}f,
\end{equation}
for quantities \(f\) defined uniquely on \(\Gamma\) which satisfy \eqref{eq:null_extension} and \eqref{eq:null_extension_vector}. 

We consider now the outward unit normal as a function defined in the whole space-time domain by means of \eqref{eq:normal_F}. Applying \eqref{eq:distro_product_rule} and \eqref{eq:grad_Xk} to \(\frac{\partial \nabla X_{k}}{\partial t}\), one obtains
\begin{equation}\label{eq:grad_Xk_t}
\frac{\partial\nabla X_k}{\partial t} = -\frac{\partial\left[\delta(\Gamma)\mathbf{n}_k	\right]}{\partial t} = -\frac{\partial\delta(\Gamma)}{\partial t}\mathbf{n}_k - \frac{\partial\mathbf{n}_k}{\partial t}\delta(\Gamma).  
\end{equation}
If we substitute \eqref{eq:grad_Xk_t} into \eqref{eq:grad_Xk_evolution} and we compute the scalar product by \(\mathbf{n}_{k}\) we obtain
\begin{equation}\label{eq:topological_proof}
-\frac{\partial\delta(\Gamma)}{\partial t} - \frac{\partial\mathbf{n}_k}{\partial t}\delta(\Gamma)\cdot\mathbf{n}_k  -\dive\left[\left(\mathbf{v}_I\cdot\mathbf{n}_k\right)\mathbf{n}_k\delta(\Gamma)\right] + \left(\mathbf{v}_I\cdot\mathbf{n}_k\right)\left(\dive\mathbf{n}_k\right)\delta(\Gamma) = 0.
\end{equation}
Since \(\mathbf{n}_{k} \cdot \mathbf{n}_{k} = 1\), one can verify that \(\frac{\partial\mathbf{n}_{k}}{\partial t} \cdot \mathbf{n}_{k} = 0\) and, therefore, \eqref{eq:topological_proof} reduces to
\begin{equation}\label{eq:delta_ev_1}
\frac{\partial\delta(\Gamma)}{\partial t} + \dive\left[\left(\mathbf{v}_I\cdot\mathbf{n}_k\right)\mathbf{n}_k\delta(\Gamma)\right] - \left(\mathbf{v}_I\cdot\mathbf{n}_k\right)\left(\dive\mathbf{n}_k\right)\delta(\Gamma) = 0.    
\end{equation}
This is a well known relation and it is described in several contributions \cite{drew:1990, junqua-moullet:2003, lhuillier:2000, marle:1982, morel:2007}. In \eqref{eq:delta_ev_1}, the term \(\dive\mathbf{n}_{k}\) is directly related to curvature effects. Indeed, the following relation holds \cite{drew:1999, weatherburn:1930, morel:2007}:
\begin{equation}\label{eq:mean_curvature_def}
H = -\frac{1}{2}\dive\mathbf{n}_{k},
\end{equation}
where \(H\) denotes the mean curvature. Equation \eqref{eq:delta_ev_1} can be rewritten in other forms. First of all, since \(\dive\mathbf{n}_{k} = \dives\mathbf{n}_{k}\) \cite{nadim:1996}, we immediately obtain
\begin{equation}\label{eq:delta_ev_2}
\frac{\partial\delta(\Gamma)}{\partial t} + \dive\left[\left(\mathbf{v}_{I} \cdot \mathbf{n}_{k}\right)\mathbf{n}_{k}\delta(\Gamma)\right] - \left(\mathbf{v}_{I} \cdot \mathbf{n}_{k}\right)\left(\dives\mathbf{n}_{k}\right)\delta(\Gamma) = 0.    
\end{equation}
Moreover, \eqref{eq:delta_ev_1} is equivalent to
\begin{equation}\label{eq:delta_ev_3}
\frac{\partial\delta(\Gamma)}{\partial t} + \left(\mathbf{v}_{I} \cdot \mathbf{n}_{k}\right)\mathbf{n}_{k} \cdot \nabla\delta(\Gamma) = -\delta(\Gamma)\nabla\left(\mathbf{v}_{I} \cdot \mathbf{n}_{k}\right)\cdot\mathbf{n}_{k},
\end{equation}
a relation present in \cite{junqua-moullet:2003}. Indeed, the following relation holds:
\begin{eqnarray}\label{eq:delta_ev_3_proof}
\dive\left[\left(\mathbf{v}_{I} \cdot \mathbf{n}_{k}\right)\mathbf{n}_{k}\delta(\Gamma)\right] &=& \left(\mathbf{v}_{I} \cdot \mathbf{n}_{k}\right)\mathbf{n}_{k} \cdot \nabla\delta(\Gamma) + \dive\left[\left(\mathbf{v}_{I} \cdot \mathbf{n}_{k}\right)\mathbf{n}_{k}\right]\delta(\Gamma) \\
&=& \left(\mathbf{v}_{I} \cdot \mathbf{n}_{k}\right)\mathbf{n}_{k} \cdot \nabla\delta(\Gamma) + \left(\mathbf{v}_{I} \cdot \mathbf{n}_{k}\right)\dive\mathbf{n}_{k} \delta(\Gamma) \nonumber \\
&+& \nabla\left(\mathbf{v}_{I} \cdot \mathbf{n}_{k}\right) \cdot \mathbf{n}_{k} \delta(\Gamma). \nonumber	
\end{eqnarray}
If we substitute \eqref{eq:delta_ev_3_proof} into \eqref{eq:delta_ev_1}, we recover \eqref{eq:delta_ev_3}. The relations \eqref{eq:delta_ev_1}, \eqref{eq:delta_ev_2} and \eqref{eq:delta_ev_3} contain only the normal velocity \(\left(\mathbf{v}_{I} \cdot \mathbf{n}_{k}\right)\mathbf{n}_{k}\). To rewrite them in order to make the complete interfacial velocity \(\mathbf{v}_{I}\) appear, we first define the tangential velocity \(\mathbf{v}_{I_t}\) as
\begin{equation}
\mathbf{v}_{I_t} = \mathbf{v}_{I} - \left(\mathbf{v}_{I} \cdot \mathbf{n}_{k}\right)\mathbf{n}_{k} = \left(\mathbf{I} - \mathbf{n}_{k} \otimes \mathbf{n}_{k}\right)\mathbf{v}_{I}.
\end{equation}
Adding and subtracting \(\dive\left[\mathbf{v}_{I_t}\delta(\Gamma)\right]\) to \eqref{eq:delta_ev_1}, we obtain the following relation:
\begin{equation}\label{eq:delta_ev_4}
\frac{\partial\delta(\Gamma)}{\partial t} + \dive\left[\mathbf{v}_{I}\delta(\Gamma)\right] - \left(\mathbf{v}_{I} \cdot \mathbf{n}_{k}\right)\left(\dive\mathbf{n}_{k}\right)\delta(\Gamma) - \dive\left[\mathbf{v}_{I_t}\delta(\Gamma)\right] = 0.    
\end{equation}
It can be proven \cite{marle:1982} that
\begin{equation}\label{eq:dive_vIt}
\dive\left[\mathbf{v}_{I_t}\delta(\Gamma)\right] = \delta(\Gamma)\dives\mathbf{v}_{I_t}. 
\end{equation}
We now propose a proof of relation \eqref{eq:dive_vIt}, which is valid considering \(\mathbf{v}_{I_{t}}\) defined in the whole space-time domain \(\Omega \times \left(0, T_{f}\right]\). First of all, since
\begin{equation}
\dive\left[\mathbf{v}_{I_t}\delta(\Gamma)\right] = \delta(\Gamma)\dive\mathbf{v}_{I_{t}} + \mathbf{v}_{I_{t}} \cdot \nabla \delta(\Gamma),
\end{equation} 
we need to analyze \(\nabla \delta(\Gamma)\). Thanks to relation \eqref{eq:grad_Xk}, we obtain
\begin{eqnarray}\label{eq:dive_vIt_proof_1}
\nabla\left(\nabla X_{k}\right) &=& -\nabla\left(\mathbf{n}_{k}\delta(\Gamma)\right) = -\left(\nabla\mathbf{n}_{k}\right)\delta(\Gamma) - \mathbf{n}_{k} \otimes \nabla\delta(\Gamma) \\
\nabla\left(\nabla X_{k}\right)^{T} &=& -\nabla\left(\mathbf{n}_{k}\delta(\Gamma)\right)^{T} = -\left(\nabla\mathbf{n}_{k}\right)^{T}\delta(\Gamma) - \nabla\delta(\Gamma) \otimes \mathbf{n}_{k}. 
\end{eqnarray}
If we multiply the previous relations by \(\mathbf{n}_{k}\), we get
\begin{eqnarray}\label{eq:dive_vIt_proof_2}
\nabla\left(\nabla X_{k}\right)\mathbf{n}_{k} &=& -\left(\nabla\mathbf{n}_{k}\right)\mathbf{n}_{k}\delta(\Gamma) - \left(\mathbf{n}_{k} \otimes \nabla\delta(\Gamma)\right)\mathbf{n}_{k} \nonumber \\
&=& -\left(\nabla\mathbf{n}_{k}\right)\mathbf{n}_{k}\delta(\Gamma) - \left(\nabla\delta(\Gamma) \cdot \mathbf{n}_{k}\right) \cdot \mathbf{n}_{k} \\
\nabla\left(\nabla X_{k}\right)^{T}\mathbf{n}_{k} &=& -\left(\nabla\mathbf{n}_{k}\right)^{T}\mathbf{n}_{k}\delta(\Gamma) - \left(\nabla\delta(\Gamma) \otimes \mathbf{n}_{k}\right)\mathbf{n}_{k} = -\nabla\delta(\Gamma), 
\end{eqnarray}
where we exploited \(\left(\nabla\mathbf{n}_{k}\right)^{T}\mathbf{n}_{k} = \mathbf{0}\). Since \(\left[\nabla\left(\nabla X_{k}\right)\right] = \left[\nabla\left(\nabla X_{k}\right)\right]^{T}\), we obtain
\begin{equation}\label{eq:grad_delta}
\nabla\delta(\Gamma) = \left(\nabla\mathbf{n}_{k}\right)\mathbf{n}_{k}\delta(\Gamma) + \left(\nabla\delta(\Gamma) \cdot \mathbf{n}_{k}\right)\mathbf{n}_{k}.
\end{equation}
Equation \eqref{eq:grad_delta} generalizes the relation for \(\nabla\delta(\Gamma)\) derived in \cite{estrada:1980}, which is valid when \(\mathbf{n}_{k}\) is defined only on \(\Gamma\). Hence, we get
\begin{eqnarray}\label{eq:dive_vIt_proof_3}
\dive\left[\mathbf{v}_{I_t}\delta(\Gamma)\right] &=& \delta(\Gamma)\dive\mathbf{v}_{I_{t}} + \mathbf{v}_{I_{t}} \cdot \nabla \delta(\Gamma) \nonumber \\
&=& \delta(\Gamma)\dive\mathbf{v}_{I_{t}} + \mathbf{v}_{I_{t}} \cdot \left[\left(\nabla\mathbf{n}_{k}\right)\mathbf{n}_{k}\delta(\Gamma) + \left(\nabla\delta(\Gamma) \cdot \mathbf{n}_{k}\right)\mathbf{n}_{k}\right].
\end{eqnarray}
Since \(\mathbf{v}_{I_{t}} \cdot \mathbf{n}_{k} = 0\) and \(\nabla\left(\mathbf{v}_{I_{t}} \cdot \mathbf{n}_{k}\right) = \left(\nabla\mathbf{v}_{I_{t}}\right)^{T}\mathbf{n}_{k} + \left(\nabla\mathbf{n}_{k}\right)^{T}\mathbf{v}_{I_{t}} = \mathbf{0}\), we obtain from \eqref{eq:dive_vIt_proof_3}
\begin{equation}\label{eq:dives_vIt_fin}
\dive\left[\mathbf{v}_{I_t}\delta(\Gamma)\right] = \delta(\Gamma)\dive\mathbf{v}_{I_{t}} - \left(\nabla\mathbf{v}_{I_{t}}\right)^{T}\mathbf{n}_{k} \cdot \mathbf{n}_{k}\delta(\Gamma) = \delta(\Gamma)\dives\mathbf{v}_{I_{t}}.
\end{equation}
If we now substitute \eqref{eq:dives_vIt_fin} into \eqref{eq:delta_ev_4}, we get
\begin{equation}\label{eq:delta_ev_5}
\frac{\partial\delta(\Gamma)}{\partial t} + \dive\left[\mathbf{v}_{I}\delta(\Gamma)\right] - \left(\mathbf{v}_{I} \cdot \mathbf{n}_{k}\right)\left(\dive\mathbf{n}_{k}\right)\delta(\Gamma) - \delta(\Gamma)\dives\mathbf{v}_{I_t} = 0
\end{equation}
It can be also shown \cite{morel:2007} that
\begin{equation}\label{eq:dives_vI}
\dives\mathbf{v}_{I} = \dives\mathbf{v}_{I_{t}} + \left(\mathbf{v}_{I} \cdot \mathbf{n}_{k}\right)\dive\mathbf{n}_{k}.
\end{equation}
Exploiting this relation in \eqref{eq:delta_ev_5} leads to the following equation \cite{junqua-moullet:2003, lhuillier:2003, morel:2007}:
\begin{equation}\label{eq:delta_ev_6}
\frac{\partial\delta(\Gamma)}{\partial t} + \dive\left[\mathbf{v}_I\delta(\Gamma)\right] = \delta(\Gamma)\dives\mathbf{v}_I.
\end{equation}
We can rewrite the term \(\dive\left[\mathbf{v}_{I}\delta(\Gamma)\right]\) present in \eqref{eq:delta_ev_6} as
\begin{equation}
\dive\left[\mathbf{v}_{I}\delta(\Gamma)\right] = \mathbf{v}_{I} \cdot \nabla\delta(\Gamma) + \delta(\Gamma)\dive\mathbf{v}_{I}
\end{equation}
and, noticing that \(\dives\mathbf{v}_{I} = \dive\mathbf{v}_{I} - \left[\left(\nabla\mathbf{v}_{I}\right)\mathbf{n}_{k}\right] \cdot \mathbf{n}_{k}\), we get
\begin{equation}\label{eq:delta_ev_7_tmp}
\frac{\partial\delta(\Gamma)}{\partial t} + \mathbf{v}_{I} \cdot \nabla\delta(\Gamma) = -\delta(\Gamma)\left[\left(\nabla\mathbf{v}_{I}\right)\mathbf{n}_{k}\right] \cdot \mathbf{n}_{k}
\end{equation}
or, equivalently, 
\begin{equation}\label{eq:delta_ev_7}
\frac{\partial\delta(\Gamma)}{\partial t} + \mathbf{v}_{I} \cdot \nabla\delta(\Gamma) = -\delta(\Gamma)\mathbf{n}_{k} \otimes \mathbf{n}_{k} : \nabla\mathbf{v}_{I},
\end{equation}
a relation which has been derived in \cite{lhuillier:2003}. As discussed in \cite{junqua-moullet:2003}, \eqref{eq:delta_ev_7} is equivalent to
\begin{equation}\label{eq:delta_ev_8}
\frac{\partial\delta(\Gamma)}{\partial t} + \left(\mathbf{v}_{I} \cdot \mathbf{n}_{k}\right)\mathbf{n}_{k} \cdot \nabla\delta(\Gamma) = -\delta(\Gamma)\nabla\left(\mathbf{v}_{I} \cdot \mathbf{n}_{k}\right) \cdot \mathbf{n}_{k}.
\end{equation}
One can also notice that
\begin{equation}
\left[\left(\nabla\mathbf{v}_{I}\right)\mathbf{n}_{k}\right] \cdot \mathbf{n}_{k} = \left[\dive\left(\mathbf{v}_{I} \otimes \mathbf{n}_{k}\right)\right] \cdot \mathbf{n}_{k} - \left(\dive\mathbf{n}_{k}\right)\left(\mathbf{v}_{I} \cdot \mathbf{n}_{k}\right).
\end{equation}
Hence, we can rewrite \eqref{eq:delta_ev_7_tmp} as follows:
\begin{equation}\label{eq:delta_ev_9}
\frac{\partial\delta(\Gamma)}{\partial t} + \mathbf{v}_{I} \cdot \nabla\delta(\Gamma) = \delta(\Gamma)\left(\dive\mathbf{n}_{k}\right)\left(\mathbf{v}_{I} \cdot \mathbf{n}_{k}\right) - \delta(\Gamma)\left[\dive\left(\mathbf{v}_{I} \otimes \mathbf{n}_{k}\right)\right] \cdot \mathbf{n}_{k}.
\end{equation}
Relation \eqref{eq:delta_ev_9} is particularly interesting because it provides an evolution equation for \(\delta(\Gamma)\) in which the complete interfacial velocity is the advecting field and a quantity related to the curvature, i.e. \(\dive\mathbf{n}_{k}\), appears as a forcing term. To the best of our knowledge, this novel formulation is not present in the literature we have reviewed and is presented here for the first time. Analogously, notice that
\begin{equation}
\left[\left(\nabla\mathbf{v}_{I}\right)\mathbf{n}_{k}\right] \cdot \mathbf{n}_{k} = \nabla\left(\mathbf{v}_{I} \cdot \mathbf{n}_{k}\right) \cdot \mathbf{n}_{k} - \dive\left(\mathbf{n}_{k} \otimes \mathbf{n}_{k}\right) \cdot \mathbf{v}_{I} + \left(\dive\mathbf{n}_{k}\right)\left(\mathbf{v}_{I} \cdot \mathbf{n}_{k}\right),
\end{equation}
so as to obtain
\begin{eqnarray}\label{eq:delta_ev_10}
&&\frac{\partial\delta(\Gamma)}{\partial t} + \mathbf{v}_{I} \cdot \nabla\delta(\Gamma) = \nonumber \\ &&-\delta(\Gamma)\left(\dive\mathbf{n}_{k}\right)\left(\mathbf{v}_{I} \cdot \mathbf{n}_{k}\right) -\delta(\Gamma)\nabla\left(\mathbf{v}_{I} \cdot \mathbf{n}_{k}\right) \cdot \mathbf{n}_{k} + \delta(\Gamma)\left[\dive\left(\mathbf{n}_{k} \otimes \mathbf{n}_{k}\right)\right] \cdot \mathbf{v}_{I}.
\end{eqnarray}
It is worthwhile to recall once more that, in all the previous relations, we have considered the outward unit normal vector and the interfacial velocity as variables already defined in the whole space-time domain \(\Omega \times \left(0, T_{f}\right]\). Relations \eqref{eq:delta_ev_1}, \eqref{eq:delta_ev_2}, \eqref{eq:delta_ev_3}, \eqref{eq:delta_ev_4}, \eqref{eq:delta_ev_5}, \eqref{eq:delta_ev_6}, \eqref{eq:delta_ev_7}, \eqref{eq:delta_ev_8}, \eqref{eq:delta_ev_9} and \eqref{eq:delta_ev_10} are valid also by considering the outward unit normal vector and the interfacial velocity as functions uniquely defined on the interface \(\Gamma\) and then analyzing their extension, that, with a slight abuse of notation, we still denote by \(\mathbf{n}_{k}\) and \(\mathbf{v}_{I}\). However, in this situation, relation \eqref{eq:null_extension_vector} allows to consider much simpler forms for the evolution equation of the interfacial area density. As evident from \eqref{eq:grad_delta} and reported in \cite{estrada:1980, estrada:1985}, the following relation holds:
\begin{equation}\label{eq:grad_delta_KE}
\nabla\delta(\Gamma) = \left[\nabla\delta(\Gamma) \cdot \mathbf{n}_{k}\right]\mathbf{n}_{k}.
\end{equation}
Indeed, since the extension of \(\mathbf{n}_{k}\) satisfies property \eqref{eq:null_extension_vector}, the term \(\left(\nabla\mathbf{n}_{k}\right)\mathbf{n}_{k}\delta(\Gamma)\) is null and, therefore, \eqref{eq:grad_delta} reduces to \eqref{eq:grad_delta_KE}. Moreover, thanks to \eqref{eq:distro_product_rule_extension}, we can rewrite the term \(\dive\left[\mathbf{v}_{I}\delta(\Gamma)\right]\) present in \eqref{eq:delta_ev_6} as
\begin{equation}
\dive\left[\mathbf{v}_{I}\delta(\Gamma)\right] = \mathbf{v}_{I} \cdot \nabla\delta(\Gamma) + \delta(\Gamma)\dives\mathbf{v}_{I}
\end{equation}
so as to obtain from \eqref{eq:delta_ev_6}
\begin{equation}\label{eq:delta_ev_11}
\frac{\partial\delta(\Gamma)}{\partial t} + \mathbf{v}_{I} \cdot \nabla\delta(\Gamma) = 0
\end{equation}
or, substituting \eqref{eq:grad_delta_KE} in \eqref{eq:delta_ev_11},
\begin{equation}\label{eq:delta_ev_12}
\frac{\partial\delta(\Gamma)}{\partial t} + \left(\mathbf{v}_{I} \cdot \mathbf{n}_{k}\right)\mathbf{n}_{k} \cdot \nabla\delta(\Gamma) = 0,
\end{equation}
a relation present in \cite{delhaye:2001}. Equation \eqref{eq:delta_ev_11} is clearly different from \eqref{eq:delta_ev_7}. However, under the assumption that the fields involved depend only on the surface coordinates, the following relation holds \cite{drew:1990, junqua-moullet:2003}: 
\begin{equation}
\delta(\Gamma)\left[\nabla\left(\mathbf{v}_{I}\right)\mathbf{n}_{k}\right] \cdot \mathbf{n}_{k} = 0
\end{equation}
and, therefore, \eqref{eq:delta_ev_7} reduces to \eqref{eq:delta_ev_11}. It is important to notice that relations \eqref{eq:delta_ev_11} and \eqref{eq:delta_ev_12} are rigorously valid and are equivalent to \eqref{eq:delta_ev_1}, \eqref{eq:delta_ev_2}, \eqref{eq:delta_ev_3}, \eqref{eq:delta_ev_4}, \eqref{eq:delta_ev_5}, \eqref{eq:delta_ev_6}, \eqref{eq:delta_ev_7}, \eqref{eq:delta_ev_8}, \eqref{eq:delta_ev_9} and \eqref{eq:delta_ev_10}, if only if one analyzes variables defined uniquely on the surface and then considers an extension satisfying the property \eqref{eq:null_extension_vector}. This is essential when performing numerical approximation. Indeed, since it is not possible to work directly with distributions in a numerical framework and it is often advantageous using equations that are valid in the whole space-time domain rather than solving PDEs on surface, relations such as \eqref{eq:delta_ev_11} are not directly applicable for numerical simulations, as we will see in Section \ref{sec:numerical}. Moreover, even employing regular functions, it is not possible to guarantee in general that \eqref{eq:null_extension} and \eqref{eq:null_extension_vector} are satisfied, from which the equivalence between \eqref{eq:delta_ev_11}, \eqref{eq:delta_ev_12} and all the other relations is established.

As pointed out at the beginning of the Section, all the previous relations are only valid under the assumption that the surface \(\Gamma\) is closed without contact lines. In case this hypothesis is not valid, we need to add an extra term which takes into account contact lines \cite{junqua-moullet:2003, marle:1982}:
\begin{equation}\label{eq:delta_ev_contact_lines}
\frac{\partial\delta(\Gamma)}{\partial t} + \dive\left[\left(\mathbf{v}_{I} \cdot \mathbf{n}_{k}\right)\mathbf{n}_{k}\delta(\Gamma)\right] = \left(\mathbf{v}_I\cdot\mathbf{n}_k\right)\left(\dive\mathbf{n}_k\right)\delta(\Gamma) - \delta(\Delta)\mathbf{v}_{\Delta}\cdot \mathbf{t}_{k}.
\end{equation}
Here, \(\delta(\Delta)\) is the Dirac delta distribution with support on the contact lines, \(\mathbf{v}_{\Delta}\) represents the velocity field of the contact lines and \(\mathbf{t}_{k}\) is the unit vector tangent to the interface and directed in outward normal direction with respect to the contact lines. \\
Evolution equations for higher order statistics such as curvatures or unit normal vector can then be considered and will be analyzed in Section \ref{sec:normal_curvature}.

\section{Evolution equations for the unit normal vector and for the mean curvature}
\label{sec:normal_curvature} \indent 

In this Section, we provide an overview of models for the evolution of geometric features of interfaces separating the two phases in two-phase flows, such as the unit normal vector and the mean curvature. We follow the analyses in \cite{drew:1990, drew:1999}, but we do not restrict ourselves \textit{a priori} to situations where the tangential interfacial velocity is equal to zero as in those references. Considering the complete interfacial velocity as advecting field can be important for those applications where the interface is not well resolved and, therefore, we cannot rely on the direct computation of the unit normal vector from gradient of the volume fraction. For the sake of simplicity in the notation, we will denote the unit normal vector by \(\mathbf{n}\) and we will consider the definition \(\mathbf{n} = \frac{\nabla F}{\left|\nabla F\right|}\). As already discussed in Section \ref{sec:interface_area}, relation \eqref{eq:normal_F} defines the normal vector for the whole space-time domain \(\Omega_{T} = \Omega \times \left(0, T_{f}\right]\). We analyze now the time evolution of the normal vector \(\mathbf{n}\). After some manipulations, we obtain
\begin{eqnarray}\label{eq:normal_evolution_base}
\frac{\partial\mathbf{n}}{\partial t} &=& \frac{1}{\left|\nabla F\right|}\frac{\partial\nabla F}{\partial t} - \frac{1}{\left|\nabla F\right|^3}\left(\nabla F \cdot \frac{\partial\nabla F}{\partial t}\right)\nabla F \nonumber \\
&=& \frac{1}{\left|\nabla F\right|}\left(\mathbf{I} - \mathbf{n} \otimes \mathbf{n}\right)\frac{\partial\nabla F}{\partial t}.
\end{eqnarray}
Taking the gradient of \eqref{eq:F_evolution}, it follows
\begin{equation}\label{eq:normal_evolution_proof}
\frac{\partial\nabla F}{\partial t} = -\left(\nabla\mathbf{v}_{I}\right)^{T}\nabla F - \left[\nabla\left(\nabla F\right)\right]^{T}\mathbf{v}_{I}.
\end{equation} 
Moreover, after some manipulations, it can be shown that
\begin{equation}\label{eq:grad_nk}
\nabla\mathbf{n} = \frac{1}{\left|\nabla F\right|}\left[\nabla\left(\nabla F\right) - \mathbf{n} \otimes \nabla\left(\nabla F\right)^{T}\mathbf{n}\right] = \frac{1}{\left|\nabla F\right|}\left(\mathbf{I} - \mathbf{n} \otimes \mathbf{n}\right)\nabla\left(\nabla F\right).
\end{equation} 
If we assume that \(F\) is sufficiently regular, thanks to the Schwarz theorem, \(\left[\nabla\left(\nabla F\right)\right]^{T} = \nabla\left(\nabla F\right)\) and substituting \eqref{eq:grad_nk} into \eqref{eq:normal_evolution_proof} we obtain the following relation:
\begin{equation}\label{eq:normal_evolution_proof_2}
\frac{\partial\nabla F}{\partial t} = -\left(\nabla\mathbf{v}_{I}\right)^{T}\nabla F - \left|\nabla F\right|\left[\left(\mathbf{I} - \mathbf{n} \otimes \mathbf{n}\right)^{-1}\left(\nabla\mathbf{n}\right)\right]\mathbf{v}_{I}.
\end{equation}
Finally, substituting \eqref{eq:normal_evolution_proof_2} into \eqref{eq:normal_evolution_base}, we obtain
\begin{equation}\label{eq:normal_evolution}
\frac{\partial\mathbf{n}}{\partial t} + \left(\nabla\mathbf{n}\right)\mathbf{v}_{I} = \left(\mathbf{n} \otimes \mathbf{n} - \mathbf{I}\right)\left(\nabla\mathbf{v}_{I}\right)^{T}\mathbf{n}
\end{equation}
or equivalently
\begin{equation}
\frac{d \mathbf{n}}{d t} = \left(\mathbf{n} \otimes \mathbf{n} - \mathbf{I}\right)\left(\nabla\mathbf{v}_{I}\right)^{T}\mathbf{n} = \left[\left(\mathbf{n} \otimes \mathbf{n}\right) : \nabla\mathbf{v}_{I}\right]\mathbf{n} - \left(\nabla\mathbf{v}_{I}\right)^{T}\mathbf{n},
\end{equation}
a relation derived in \cite{candel:1990, lhuillier:2003}. On the other hand, substituting \eqref{eq:normal_F} into \eqref{eq:F_evolution}, we obtain the following relation:
\begin{equation}\label{eq:normal_evolution_bis_proof}
\frac{\partial F}{\partial t} + \left(\mathbf{v}_{I} \cdot \mathbf{n}\right)\left|\nabla F\right| = 0.
\end{equation}
Taking the gradient of \eqref{eq:normal_evolution_bis_proof}, we get
\begin{equation}\label{eq:normal_evolution_bis_proof_b}
\frac{\partial\nabla F}{\partial t} + \nabla\left(\mathbf{v}_{I} \cdot \mathbf{n}\right)\left|\nabla F\right| + \left(\mathbf{v}_{I} \cdot \mathbf{n}\right)\nabla\left(\left|\nabla F\right|\right) = \mathbf{0}.
\end{equation}
Since \(\nabla\left(\left|\nabla F\right|\right) = \left[\nabla\left(\nabla F\right)\right]^{T}\mathbf{n} = \left[\nabla\left(\nabla F\right)\right]\mathbf{n}\), we obtain from \eqref{eq:normal_evolution_bis_proof_b}
\begin{equation}\label{eq:normal_evolution_bis_proof_c}
\frac{\partial\nabla F}{\partial t} = -\nabla\left(\mathbf{v}_{I} \cdot \mathbf{n}\right)\left|\nabla F\right| - \left(\mathbf{v}_{I} \cdot \mathbf{n}\right)\left[\nabla\left(\nabla F\right)\right]\mathbf{n}.
\end{equation}
Substituting \eqref{eq:grad_nk} into \eqref{eq:normal_evolution_bis_proof_c}, we obtain
\begin{equation}\label{eq:normal_evolution_bis_proof_d}
\frac{\partial\nabla F}{\partial t} = -\nabla\left(\mathbf{v}_{I} \cdot \mathbf{n}\right)\left|\nabla F\right| - \left(\mathbf{v}_{I} \cdot \mathbf{n}\right)\left|\nabla F\right|\left(\mathbf{I} - \mathbf{n} \otimes \mathbf{n}\right)^{-1}\left(\nabla\mathbf{n}\right)\mathbf{n}.
\end{equation}
If we employ the previous relation into \eqref{eq:normal_evolution_base}, we obtain
\begin{equation}\label{eq:normal_evolution_bis}
\frac{\partial\mathbf{n}}{\partial t} + \left(\mathbf{v}_{I} \cdot \mathbf{n}\right)\left(\nabla\mathbf{n}\right)\mathbf{n} = \left(\mathbf{n} \otimes \mathbf{n} - \mathbf{I}\right)\nabla\left(\mathbf{v}_{I} \cdot \mathbf{n}\right)
\end{equation}
or, equivalently, thanks to \eqref{eq:time_surface_derivative_product} and \eqref{eq:space_surface_derivative_product}
\begin{equation}
\frac{\partial_{s}\mathbf{n}}{\partial t} = -\nabla_{s}\left(\mathbf{v}_{I} \cdot \mathbf{n}\right).
\end{equation}
Notice that equation \eqref{eq:normal_evolution_bis} can be directly obtained from \eqref{eq:normal_evolution} considering only the normal component of the interfacial velocity, namely taking \(\mathbf{v}_{I} = \left(\mathbf{v}_{I} \cdot \mathbf{n}\right)\mathbf{n}\). This further confirms the consideration in Section \ref{sec:interface_area} that two interfacial velocity fields with the same normal component yield the same Lagrangian derivative. Notice also that, as discussed in Section \ref{sec:interface_area}, if one considers the unit normal vector and the interfacial velocity as defined uniquely on the interface and analyzes their extension, \(\left(\nabla\mathbf{n}\right)\mathbf{n} = \mathbf{0}\) and \(\nabla\left(\mathbf{v}_{I} \cdot \mathbf{n}\right) \cdot \mathbf{n} = 0\). Hence, \eqref{eq:normal_evolution_bis} reduces to 
\begin{equation}
\frac{\partial\mathbf{n}}{\partial t} = -\nabla\left(\mathbf{v}_{I} \cdot \mathbf{n}\right),
\end{equation}
a relation present in \cite{drew:1999}. Moreover, if \(\left|\nabla F\right|\) is constant, the second order tensor \(\nabla\mathbf{n}\) is symmetric and, therefore, in this situation, \eqref{eq:normal_evolution_bis} reduces to
\begin{equation}
\frac{\partial\mathbf{n}}{\partial t} = \left(\mathbf{n} \otimes \mathbf{n} - \mathbf{I}\right)\nabla\left(\mathbf{v}_{I} \cdot \mathbf{n}\right).
\end{equation}
By comparing \eqref{eq:normal_evolution} and \eqref{eq:normal_evolution_bis}, we obtain the following relation:
\begin{equation}\label{eq:zero_term_comparison_normal_evolution}
\left(\nabla\mathbf{n}\right)\mathbf{v}_{I_{t}} + \left(\mathbf{n} \otimes \mathbf{n} - \mathbf{I}\right)\left(\nabla\mathbf{n}\right)^{T}\mathbf{v}_{I} = \left(\nabla\mathbf{n}\right)\mathbf{v}_{I_{t}} + \left(\mathbf{n} \otimes \mathbf{n} - \mathbf{I}\right)\left(\nabla\mathbf{n}\right)^{T}\mathbf{v}_{I_{t}} = \mathbf{0}. 
\end{equation}
It is also of interest to study the behaviour of the material derivative of the normal vector following the surface. The convective derivative following the surface can be simplified as follows:
\begin{equation}
\frac{d_{s}\mathbf{n}}{dt} = \frac{\partial_{s}\mathbf{n}}{\partial t} + \left(\nabla_{s}\mathbf{n}\right)\left[\mathbf{v}_{I_t} + \left(\mathbf{v}_{I} \cdot \mathbf{n}\right)\mathbf{n}\right] = \frac{\partial_{s}\mathbf{n}}{\partial t} + \left(\nabla_{s}\mathbf{n}\right)\mathbf{v}_{I_t},
\end{equation}
since \(\left(\nabla_{s}\mathbf{n}\right)\mathbf{n} = \mathbf{0}\). This is a direct consequence of the relation
\begin{equation}
\nabla_{s}\mathbf{n} = \nabla\mathbf{n} - \left(\nabla\mathbf{n}\right)\mathbf{n} \otimes \mathbf{n} =  \nabla\mathbf{n} - \nabla\mathbf{n}\left(\mathbf{n} \otimes \mathbf{n}\right) = \nabla\mathbf{n}\left(\mathbf{I} - \mathbf{n} \otimes \mathbf{n}\right).
\end{equation}
Furthermore, thanks to \eqref{eq:space_surface_derivative_product}, the following identity holds:
\begin{eqnarray}
\frac{d_{s}\mathbf{n}}{dt} &=& \frac{\partial_{s}\mathbf{n}}{\partial t} + \left(\nabla_{s}\mathbf{n}\right)\mathbf{v}_{I} = \frac{\partial\mathbf{n}}{\partial t} + \left(\mathbf{v}_{I} \cdot \mathbf{n}\right)\left(\nabla\mathbf{n}\right)\mathbf{n} + \left(\nabla\mathbf{n}\right)\left(\mathbf{I} - \mathbf{n} \otimes \mathbf{n}\right)\mathbf{v}_{I} \nonumber \\
&=& \frac{\partial\mathbf{n}}{\partial t} + \left(\mathbf{v}_{I} \cdot \mathbf{n}\right)\left(\nabla\mathbf{n}\right)\mathbf{n} + \left(\nabla\mathbf{n}\right)\mathbf{v}_{I} - \left(\mathbf{v}_{I}\cdot\mathbf{n}\right)\left(\nabla\mathbf{n}\right)\mathbf{n} \nonumber \\
&=& \frac{\partial\mathbf{n}}{\partial t} + \left(\nabla\mathbf{n}\right)\mathbf{v}_{I} = \frac{d\mathbf{n}}{dt}.
\end{eqnarray}
It is interesting to point out that an analogous relation holds for a generic scalar field \(f\); indeed:
\begin{eqnarray}
\frac{d_{s}f}{dt} &=& \frac{\partial_{s} f}{\partial t} + \mathbf{v}_{I} \cdot \nabla_{s}f = \frac{\partial f}{\partial t} + \left(\mathbf{v}_{I} \cdot \mathbf{n}\right)\left(\nabla f \cdot \mathbf{n}\right) + \mathbf{v}_{I} \cdot \left[\nabla f - \left(\nabla f \cdot \mathbf{n}\right)\mathbf{n}\right] \nonumber \\
&=& \frac{\partial_{s} f}{\partial t} + \mathbf{v}_{I} \cdot \nabla f + \left(\mathbf{v}_{I} \cdot \mathbf{n}\right)\left(\nabla f \cdot \mathbf{n}\right) - \left(\mathbf{v}_{I} \cdot \mathbf{n}\right)\left(\nabla f \cdot \mathbf{n}\right) \nonumber \\ 
&=& \frac{\partial f}{\partial t} + \mathbf{v}_{I} \cdot \nabla f = \frac{df}{dt}.
\end{eqnarray}

The mean curvature \(H\) is directly linked to the outward unit normal by relation \cite{drew:1999, morel:2007, weatherburn:1930}
\begin{equation}\label{eq:mean_curvature}
H = -\frac{1}{2}\dive\mathbf{n}.
\end{equation}
Taking the divergence of \eqref{eq:normal_evolution}, we derive the evolution equation for the mean curvature \eqref{eq:mean_curvature_def}. Notice that
\begin{equation}
\dive\left[\left(\nabla\mathbf{n}\right)\mathbf{v}_{I}\right] = -2\mathbf{v}_{I} \cdot \nabla H + \nabla\mathbf{n} : \left(\nabla\mathbf{v}_{I}\right)^{T}
\end{equation}
and that
\begin{eqnarray}
\dive\left[\left(\mathbf{n} \otimes \mathbf{n} - \mathbf{I}\right)\left(\nabla\mathbf{v}_{I}\right)^{T}\mathbf{n}\right] &=& -2H\left(\nabla\mathbf{v}_{I}\right)\mathbf{n} \cdot \mathbf{n} + \left(\nabla\mathbf{v}_{I}\right)^{T}\mathbf{n} \cdot \left(\nabla\mathbf{n}\right)\mathbf{n} \nonumber \\
&+& \left(\mathbf{n} \otimes \mathbf{n} - \mathbf{I}\right) : \left[\nabla\left[\left(\nabla\mathbf{v}_{I}\right)^{T}\mathbf{n}\right]\right]^{T}.
\end{eqnarray}
Hence, the evolution equation for the mean curvature reads as follows:
\begin{eqnarray}\label{eq:mean_curvature_evolution}
\frac{\partial H}{\partial t} + \mathbf{v}_{I} \cdot \nabla H &=& H\left(\nabla\mathbf{v}_{I}\right)\mathbf{n} \cdot \mathbf{n} + \frac{1}{2}\nabla\mathbf{n} : \left(\nabla\mathbf{v}_{I}\right)^{T} \nonumber \\
&-& \frac{1}{2}\left(\nabla\mathbf{v}_{I}\right)^{T}\mathbf{n} \cdot \left(\nabla\mathbf{n}\right)\mathbf{n} - \frac{1}{2}\left(\mathbf{n} \otimes \mathbf{n} - \mathbf{I}\right) : \left[\nabla\left[\left(\nabla\mathbf{v}_{I}\right)^{T}\mathbf{n}\right]\right]^{T}.
\end{eqnarray}
Starting from \eqref{eq:normal_evolution_bis}, we obtain the following relation:
\begin{eqnarray}\label{eq:mean_curvature_evolution_bis}
\frac{\partial H}{\partial t} + \left(\mathbf{v}_{I} \cdot \mathbf{n}\right)\mathbf{n} \cdot \nabla H &=& H\nabla\left(\mathbf{v}_{I} \cdot \mathbf{n}\right) \cdot \mathbf{n} - \frac{1}{2}\left(\mathbf{n} \otimes \mathbf{n} - \mathbf{I}\right) : \nabla\left[\nabla\left(\mathbf{v}_{I} \cdot \mathbf{n}\right)\right] \nonumber \\
&+& \frac{1}{2}\left(\nabla\mathbf{n}\right) : \nabla\left[\left(\mathbf{v}_{I} \cdot \mathbf{n}\right)\mathbf{n}\right]^{T} - \frac{1}{2}\left(\nabla\mathbf{n}\right)\mathbf{n} \cdot \nabla\left(\mathbf{v}_{I} \cdot \mathbf{n}\right).
\end{eqnarray}
Notice that, as already observed for \eqref{eq:normal_evolution_bis} and \eqref{eq:normal_evolution}, relation \eqref{eq:mean_curvature_evolution_bis} can be directly obtained from \eqref{eq:mean_curvature_evolution} considering only the normal component of the interfacial velocity. We recall here the relation between the Gaussian curvature \(K\) and the unit normal vector \cite{weatherburn:1930}:
\begin{equation}\label{eq:gaussian_curvature_def}
K = \frac{1}{2}\dive\left[\mathbf{n}\left(\dive\mathbf{n}\right) + \mathbf{n} \times \left(\nabla\times\mathbf{n}\right)\right].
\end{equation}
After some manipulations (see Appendix \ref{sec:mean_curvature_equations_equivalence}), it can be shown that \eqref{eq:mean_curvature_evolution_bis} reduces to
\begin{equation}\label{eq:mean_curvature_evolution_tris}
\frac{\partial_{s}H}{\partial t} = \left(2H^{2} - K\right)\left(\mathbf{v}_{I} \cdot \mathbf{n}\right) + \frac{1}{2}\dives\left[\nabla_{s}\left(\mathbf{v}_{I} \cdot \mathbf{n}\right)\right].
\end{equation}
The relation \eqref{eq:mean_curvature_evolution_tris} represents an extension of the evolution equation derived in \cite{drew:1999}, which we report here for the convenience of the reader:
\begin{equation}\label{eq:mean_curvature_evolution_drew}
\frac{\partial\left(-H\right)}{\partial t} = -\left(2H^{2} - K\right)\left(\mathbf{v}_{I} \cdot \mathbf{n}\right) - \frac{1}{2}\dive\left[\nabla\left(\mathbf{v}_{I} \cdot \mathbf{n}\right)\right]. 
\end{equation}
Notice that in \cite{drew:1999}, the definition
\begin{equation}
H = \frac{1}{2}\dive\mathbf{n}
\end{equation}
was adopted. However, for the sake of coherence with the results reported in Section \ref{sec:numerical}, we rely on \eqref{eq:mean_curvature}. The relation \eqref{eq:mean_curvature_evolution_drew} reduces to \eqref{eq:mean_curvature_evolution_tris} if all the variables are uniquely defined on the interface and one considers extensions which satisfy the properties \eqref{eq:null_extension} and \eqref{eq:null_extension_vector}. 

\section{Numerical tests}
\label{sec:numerical} \indent

In this Section, we compare numerically some of the relations presented in Section \ref{sec:interface_area}. In particular, we focus on \eqref{eq:delta_ev_7}, \eqref{eq:delta_ev_9}, and \eqref{eq:delta_ev_11}. \eqref{eq:delta_ev_7} and \eqref{eq:delta_ev_11} are clearly different and, since we cannot work with distributions when performing numerical approximations, a different behaviour is expected. For the sake of simplicity in the notation, we omit the explicit dependence on the interface for the Dirac delta \(\delta(\Gamma)\). Before proceeding to the description of relevant test cases, we outline the time and space discretization strategies chosen for \eqref{eq:delta_ev_7}, \eqref{eq:delta_ev_9}, and \eqref{eq:delta_ev_11}. First of all, we point out that the interfacial velocity \(\mathbf{v}_{I}\) and the outward unit normal vector \(\mathbf{n}\) are considered given. Indeed, they are obtained from Navier-Stokes equations coupled with a level set method, which are solved independently of \eqref{eq:delta_ev_7} and \eqref{eq:delta_ev_11}. For this purpose, we employ the implicit DG solver for incompressible two-phase flows with an artificial compressibility formulation developed in \cite{orlando:2023b} as an extension of the solver for single-phase incompressible Navier-Stokes equations presented in \cite{orlando:2022}, to which we both refer for a detailed description of the discretization scheme. The conservative level set (CLS) method \cite{olsson:2005, olsson:2007} in combination with a reinitialization procedure is employed to capture the moving interface. The time discretization is based on the L-stable two-stage TR-BDF2 method, which has been originally developed in \cite{bank:1985} as a combination of the trapezoidal rule and of the Backward Differentiation Formula method of order 2 (BDF2). We refer also to \cite{bonaventura:2017} for a complete analysis of the scheme. Let \(\Delta t\) be a discrete time step and \(t^{n} = n\Delta t, n = 0, \dots, N\), be discrete time levels. The first stage for \eqref{eq:delta_ev_11} reads as follows:
\begin{equation}\label{eq:first_stage_delta_wrong}
\frac{\delta^{n + \gamma} - \delta^{n}}{\gamma \Delta t} + \frac{1}{2}\mathbf{v}_{I}^{n+\gamma} \cdot \nabla \delta^{n+\gamma} + \frac{1}{2}\mathbf{v}_{I}^{n} \cdot \nabla\delta^{n} = 0,
\end{equation} 
where \(\gamma = 2 - \sqrt{2}\). The second stage reads as follows:
\begin{equation}\label{eq:second_stage_delta_wrong}
\frac{\delta^{n+1} - \delta^{n+\gamma}}{\left(1-\gamma\right)\Delta t} + a_{33}\mathbf{v}_{I}^{n + 1} \cdot \nabla\delta^{n+1} + a_{32}\mathbf{v}_{I}^{n+\gamma} \cdot \nabla\delta^{n+\gamma} + a_{31}\mathbf{v}_{I}^{n} \cdot \nabla\delta^{n} = 0,	
\end{equation}
where
\begin{equation}
a_{31} = \frac{1-\gamma}{2\left(2-\gamma\right)} \qquad
a_{32} = \frac{1-\gamma}{2\left(2-\gamma\right)} \qquad
a_{33} = \frac{1}{2-\gamma}.
\end{equation}
Analogously, the first stage for \eqref{eq:delta_ev_7} reads as follows:
\begin{eqnarray}\label{eq:first_stage_delta}
&&\frac{\delta^{n + \gamma} - \delta^{n}}{\gamma \Delta t} + \frac{1}{2}\mathbf{v}_{I}^{n+\gamma} \cdot \nabla \delta^{n+\gamma} + \frac{1}{2}\mathbf{v}_{I}^{n} \cdot \nabla\delta^{n} = \nonumber \\ &&-\frac{1}{2}\delta^{n+\gamma}\mathbf{n}^{n+\gamma} \otimes \mathbf{n}^{n+\gamma} : \nabla\mathbf{v}_{I}^{n+\gamma} -\frac{1}{2}\delta^{n}\mathbf{n}^{n} \otimes \mathbf{n}^{n} : \nabla\mathbf{v}_{I}^{n},
\end{eqnarray} 
whereas the second stage reads as follows:
\begin{eqnarray}\label{eq:second_stage_delta}
&&\frac{\delta^{n+1} - \delta^{n+\gamma}}{\left(1-\gamma\right)\Delta t} + a_{33}\mathbf{v}_{I}^{n+1} \cdot \nabla\delta^{n+1} + a_{32}\mathbf{v}_{I}^{n+\gamma} \cdot \nabla\delta^{n+\gamma} + a_{31}\mathbf{v}_{I}^{n} \cdot \nabla\delta^{n} = \nonumber \\
&&-a_{33}\delta^{n+1}\mathbf{n}^{n+1} \otimes \mathbf{n}^{n+1} : \nabla\mathbf{v}_{I}^{n+1} -a_{32}\delta^{n+\gamma}\mathbf{n}^{n+\gamma} \otimes \mathbf{n}^{n+\gamma} : \nabla\mathbf{v}_{I}^{n+\gamma} \nonumber \\ &&-a_{31}\delta^{n}\mathbf{n}^{n} \otimes \mathbf{n}^{n} : \nabla\mathbf{v}_{I}^{n}.
\end{eqnarray}
Finally, the first stage for \eqref{eq:delta_ev_9} reads as follows:
\begin{eqnarray}\label{eq:first_stage_delta_new}
&&\frac{\delta^{n + \gamma} - \delta^{n}}{\gamma \Delta t} + \frac{1}{2}\mathbf{v}_{I}^{n+\gamma} \cdot \nabla \delta^{n+\gamma} + \frac{1}{2}\mathbf{v}_{I}^{n} \cdot \nabla\delta^{n} = \nonumber \\ &&\frac{1}{2}\delta^{n + \gamma}\left(\dive\mathbf{n}^{n + \gamma}\right)\left(\mathbf{v}_{I}^{n + \gamma} \cdot \mathbf{n}^{n + \gamma}\right) - \frac{1}{2}\delta^{n + \gamma}\left[\dive\left(\mathbf{v}_{I}^{n + \gamma} \otimes \mathbf{n}^{n + \gamma}\right)\right] \cdot \mathbf{n}^{n + \gamma} \nonumber \\
&+&\frac{1}{2}\delta^{n}\left(\dive\mathbf{n}^{n}\right)\left(\mathbf{v}_{I}^{n} \cdot \mathbf{n}^{n}\right) - \frac{1}{2}\delta^{n}\left[\dive\left(\mathbf{v}_{I}^{n} \otimes \mathbf{n}^{n}\right)\right] \cdot \mathbf{n}^{n},
\end{eqnarray} 
whereas the second stage reads as follows:
\begin{eqnarray}\label{eq:second_stage_delta_new}
&&\frac{\delta^{n+1} - \delta^{n+\gamma}}{\left(1-\gamma\right)\Delta t} + a_{33}\mathbf{v}_{I}^{n+1} \cdot \nabla\delta^{n+1} + a_{32}\mathbf{v}_{I}^{n+\gamma} \cdot \nabla\delta^{n+\gamma} + a_{31}\mathbf{v}_{I}^{n} \cdot \nabla\delta^{n} = \nonumber \\
&&-a_{33}\delta^{n+1}\left(\dive\mathbf{n}^{n+1}\right)\left(\mathbf{v}_{I}^{n+1} \cdot \mathbf{n}^{n+1}\right) - a_{33}\delta^{n+1}\left[\dive\left(\mathbf{v}_{I}^{n+1} \otimes \mathbf{n}^{n+1}\right)\right] \cdot \mathbf{n}^{n+1} \nonumber \\
&&-a_{32}\delta^{n + \gamma}\left(\dive\mathbf{n}^{n + \gamma}\right)\left(\mathbf{v}_{I}^{n + \gamma} \cdot \mathbf{n}^{n + \gamma}\right) - a_{32}\delta^{n + \gamma}\left[\dive\left(\mathbf{v}_{I}^{n + \gamma} \otimes \mathbf{n}^{n + \gamma}\right)\right] \cdot \mathbf{n}^{n + \gamma} \nonumber \\ 
&&-a_{31}\delta^{n}\left(\dive\mathbf{n}^{n}\right)\left(\mathbf{v}_{I}^{n} \cdot \mathbf{n}^{n}\right) - a_{31}\delta^{n}\left[\dive\left(\mathbf{v}_{I}^{n} \otimes \mathbf{n}^{n}\right)\right] \cdot \mathbf{n}^{n}.
\end{eqnarray}
CLS method assumes that the level set function is a regularized Heaviside function. We recall here the relation between \(\delta(\Gamma)\) and \(\delta(F)\), namely the Dirac delta distribution with support equal to the function \(F\). As discussed in \cite{estrada:1980}, the following relation holds:
\begin{equation}
\delta(\Gamma) = \delta(F)\left|\nabla F\right|,
\end{equation}
so that we can set \(\delta^{0} = \left|\nabla\phi^{0}\right|\), with \(\phi\) denoting the level set function. For what concerns the spatial discretization, we consider discontinuous finite element approximations. The spatial discretization coincides with that described in \cite{arndt:2022} and implemented in the \textit{deal.II} library, which has been employed for the numerical simulations. We consider a decomposition of the domain \(\Omega\) into a triangulation \(\mathcal{T}_{h}\) and denote each element by \(K\). The skeleton \(\mathcal{E}\) denotes the set of all element faces and \(\mathcal{E} = \mathcal{E}^{I} \cup \mathcal{E}^{B}\), where \(\mathcal{E}^{I}\) is the subset of interior faces and \(\mathcal{E}^{B}\) is the subset of boundary faces. Suitable jump and average operators can then be defined as customary for discontinuous finite element discretizations. A face \(e \in \mathcal{E}^{I}\) shares two elements that we denote by \(K^{+}\) with outward unit normal \(\mathbf{n}_{e}^{+}\) and \(K^{-}\) with outward unit normal \(\mathbf{n}_{e}^{-}\), whereas for a face \(e \in \mathcal{E}^{B}\) we denote by \(\mathbf{n}_{e}\) the outward unit normal. For a scalar function \(\varphi\) the jump is defined as
\begin{equation}
\left[\left[\varphi\right]\right] = \varphi^{+}\mathbf{n}_{e}^{+} + \varphi^{-}\mathbf{n}_{e}^{-} \quad \text{if } e \in \mathcal{E}^{I} \qquad \left[\left[\varphi\right]\right] = \varphi\mathbf{n}_{e} \quad \text{if } e \in \mathcal{E}^{B},
\end{equation}
whereas the average is defined as
\begin{equation}
\left\{\left\{\varphi\right\}\right\} = \frac{1}{2}\left(\varphi^{+} + \varphi^{-}\right) \quad \text{if } e \in \mathcal{E}^{I} \qquad \left\{\left\{\varphi\right\}\right\} = \varphi \quad \text{if } e \in \mathcal{E}^{B}.	
\end{equation}
Similar definitions apply for a vector function \(\boldsymbol{\varphi}\):
\begin{align}
&\left[\left[\boldsymbol{\varphi}\right]\right] = \boldsymbol{\varphi}^{+} \cdot \mathbf{n}_{e}^{+} + \boldsymbol{\varphi}^{-} \cdot \mathbf{n}_{e}^{-} \quad \text{if } e \in \mathcal{E}^{I} \qquad 
\left[\left[\boldsymbol{\varphi}\right]\right] = \boldsymbol{\varphi} \cdot \mathbf{n}_{e} \quad \text{if } e \in \mathcal{E}^{B} \\
&\left\{\left\{\boldsymbol{\varphi}\right\}\right\} = \frac{1}{2}\left(\boldsymbol{\varphi}^{+} + \boldsymbol{\varphi}^{-}\right) \quad \text{if } e \in \mathcal{E}^{I} \qquad \left\{\left\{\boldsymbol{\varphi}\right\}\right\} = \boldsymbol{\varphi} \quad \text{if } e \in \mathcal{E}^{B}.
\end{align}
We now introduce the following finite element space:
\[Q_{k} = \left\{v \in L^2(\Omega) : v\rvert_{K} \in \mathbb{Q}_{k} \quad \forall K \in \mathcal{T}_{h}\right\},\] 
where \(\mathbb{Q}_{k}\) is the space of polynomials of degree \(k\) in each coordinate direction. We consider \(X_{h} = Q_{2}\) for the discretization of \(\delta\). Given these definitions, the weak formulation associated to \eqref{eq:first_stage_delta_wrong} reads as follows:
\begin{eqnarray}
&&\sum_{K \in \mathcal{T}_{h}}\int_{K} \frac{\delta^{n+\gamma}}{\gamma \Delta t}w d\Omega + \frac{1}{2}\sum_{K \in \mathcal{T}_{h}}\int_{K} \mathbf{v}_{I}^{n+\gamma} \cdot \nabla\delta^{n+\gamma} w d\Omega \nonumber \\
&+& \frac{1}{2}\sum_{e \in \mathcal{E}} \int_{e} \left\{\left\{\delta^{n+\gamma}\mathbf{v}_{I}^{n+\gamma}\right\}\right\} \cdot \left[\left[w\right]\right]d\Sigma - \frac{1}{2}\sum_{e \in \mathcal{E}} \int_{e} \left\{\left\{\mathbf{v}_{I}^{n+\gamma}\right\}\right\} \cdot \left[\left[\delta^{n+\gamma}w\right]\right]d\Sigma \nonumber \\
&+& \frac{1}{2}\sum_{e \in \mathcal{E}} \int_{e} \frac{\lambda^{n + \gamma}}{2}\left[\left[\delta^{n+\gamma}\right]\right] \cdot \left[\left[w\right]\right]d\Sigma \nonumber \\
&=&\sum_{K \in \mathcal{T}_{h}}\int_{K} \frac{\delta^{n}}{\gamma \Delta t}w d\Omega - \frac{1}{2}\sum_{K \in \mathcal{T}_{h}}\int_{K} \mathbf{v}_{I}^{n} \cdot \nabla\delta^{n} w d\Omega \\
&-& \frac{1}{2}\sum_{e \in \mathcal{E}} \int_{e} \left\{\left\{\delta^{n}\mathbf{v}_{I}^{n}\right\}\right\} \cdot \left[\left[w\right]\right]d\Sigma - \frac{1}{2}\sum_{e \in \mathcal{E}} \int_{e} \left\{\left\{\mathbf{v}_{I}^{n}\right\}\right\} \cdot \left[\left[\delta^{n}w\right]\right]d\Sigma \nonumber \\
&-& \frac{1}{2}\sum_{e \in \mathcal{E}} \int_{e} \frac{\lambda^{n}}{2}\left[\left[\delta^{n}\right]\right] \cdot \left[\left[w\right]\right]d\Sigma \qquad \forall w \in X_{h}, \nonumber
\end{eqnarray}
where
\begin{eqnarray}
\lambda^{n+\gamma} &=& \max\left(\left|\left(\mathbf{v}_{I}^{n+\gamma}\right)^{+} \cdot \mathbf{n}_{e}^{+}\right|, \left|\left(\mathbf{v}_{I}^{n+\gamma}\right)^{-} \cdot \mathbf{n}_{e}^{-}\right|\right) \\
\lambda^{n} &=& \max\left(\left|\left(\mathbf{v}_{I}^{n}\right)^{+} \cdot \mathbf{n}_{e}^{+}\right|, \left|\left(\mathbf{v}_{I}^{n}\right)^{-} \cdot \mathbf{n}_{e}^{-}\right|\right).
\end{eqnarray}
The numerical approximation of the non-conservative term is based on the approach proposed in \cite{bassi:1997} and recalled in \cite{tumolo:2015}. We recast the non-conservative term into the sum of two contributions: the first one takes into account the elementwise gradient, whereas the second one considers its jumps across the element faces. Analogously, the weak formulation for \eqref{eq:first_stage_delta} reads as follows:
\begin{eqnarray}
&&\sum_{K \in \mathcal{T}_{h}}\int_{K} \frac{\delta^{n+\gamma}}{\gamma \Delta t}w d\Omega + \frac{1}{2}\sum_{K \in \mathcal{T}_{h}}\int_{K} \mathbf{v}_{I}^{n+\gamma} \cdot \nabla\delta^{n+\gamma} w d\Omega \nonumber \\
&+& \frac{1}{2}\sum_{e \in \mathcal{E}} \int_{e} \left\{\left\{\delta^{n+\gamma}\mathbf{v}_{I}^{n+\gamma}\right\}\right\} \cdot \left[\left[w\right]\right]d\Sigma - \frac{1}{2}\sum_{e \in \mathcal{E}} \int_{e} \left\{\left\{\mathbf{v}_{I}^{n+\gamma}\right\}\right\} \cdot \left[\left[\delta^{n+\gamma}w\right]\right]d\Sigma \nonumber \\
&+& \frac{1}{2}\sum_{e \in \mathcal{E}} \int_{e} \frac{\lambda^{n + \gamma}}{2}\left[\left[\delta^{n+\gamma}\right]\right] \cdot \left[\left[w\right]\right]d\Sigma + \frac{1}{2}\sum_{K \in \mathcal{T}_{h}}\int_{K} \delta^{n+\gamma} \mathbf{n}^{n+\gamma} \otimes \mathbf{n}^{n+\gamma} : \nabla\mathbf{v}_{I}^{n+\gamma} w d\Omega \nonumber \\
&+& \frac{1}{2}\sum_{e \in \mathcal{E}} \int_{e} \left\{\left\{\left(\mathbf{v}_{I}^{n+\gamma} \cdot \mathbf{n}^{n+\gamma}\right)\mathbf{n}^{n+\gamma}\delta^{n+\gamma}\right\}\right\} \cdot \left[\left[w\right]\right]d\Sigma \nonumber \\
&-& \frac{1}{2}\sum_{e \in \mathcal{E}} \int_{e} \left\{\left\{\mathbf{n}^{n+\gamma} \otimes \mathbf{n}^{n+\gamma}\delta^{n+\gamma}\right\}\right\} : \left<\left<\mathbf{v}_{I}^{n+\gamma}w\right>\right> \nonumber \\
&=&\sum_{K \in \mathcal{T}_{h}}\int_{K} \frac{\delta^{n}}{\gamma \Delta t}w d\Omega - \frac{1}{2}\sum_{K \in \mathcal{T}_{h}}\int_{K} \mathbf{v}_{I}^{n} \cdot \nabla\delta^{n} w d\Omega \\
&-& \frac{1}{2}\sum_{e \in \mathcal{E}} \int_{e} \left\{\left\{\delta^{n}\mathbf{v}_{I}^{n}\right\}\right\} \cdot \left[\left[w\right]\right]d\Sigma - \frac{1}{2}\sum_{e \in \mathcal{E}} \int_{e} \left\{\left\{\mathbf{v}_{I}^{n}\right\}\right\} \cdot \left[\left[\delta^{n}w\right]\right]d\Sigma \nonumber \\
&-& \frac{1}{2}\sum_{e \in \mathcal{E}} \int_{e} \frac{\lambda^{n}}{2}\left[\left[\delta^{n}\right]\right] \cdot \left[\left[w\right]\right]d\Sigma - \frac{1}{2}\sum_{K \in \mathcal{T}_{h}}\int_{K} \delta^{n} \mathbf{n}^{n} \otimes \mathbf{n}^{n} : \nabla\mathbf{v}_{I}^{n} w d\Omega \nonumber \\
&-& \frac{1}{2}\sum_{e \in \mathcal{E}} \int_{e} \left\{\left\{\left(\mathbf{v}_{I}^{n} \cdot \mathbf{n}^{n}\right)\mathbf{n}^{n}\delta^{n}\right\}\right\} \cdot \left[\left[w\right]\right]d\Sigma \nonumber \\
&+& \frac{1}{2}\sum_{e \in \mathcal{E}} \int_{e} \left\{\left\{\mathbf{n}^{n} \otimes \mathbf{n}^{n}\delta^{n}\right\}\right\} : \left<\left<\mathbf{v}_{I}^{n}w\right>\right> \qquad \forall w \in X_{h}. \nonumber
\end{eqnarray}
Finally, after some manipulations, we can rewrite the weak formulation for \eqref{eq:first_stage_delta_new} as follows:
\begin{eqnarray}
&&\sum_{K \in \mathcal{T}_{h}}\int_{K} \frac{\delta^{n+\gamma}}{\gamma \Delta t}w d\Omega + \frac{1}{2}\sum_{K \in \mathcal{T}_{h}}\int_{K} \mathbf{v}_{I}^{n+\gamma} \cdot \nabla\delta^{n+\gamma} w d\Omega \nonumber \\
&+& \frac{1}{2}\sum_{e \in \mathcal{E}} \int_{e} \left\{\left\{\delta^{n+\gamma}\mathbf{v}_{I}^{n+\gamma}\right\}\right\} \cdot \left[\left[w\right]\right]d\Sigma - \frac{1}{2}\sum_{e \in \mathcal{E}} \int_{e} \left\{\left\{\mathbf{v}_{I}^{n+\gamma}\right\}\right\} \cdot \left[\left[\delta^{n+\gamma}w\right]\right]d\Sigma \nonumber \\
&+& \frac{1}{2}\sum_{e \in \mathcal{E}} \int_{e} \frac{\lambda^{n + \gamma}}{2}\left[\left[\delta^{n+\gamma}\right]\right] \cdot \left[\left[w\right]\right]d\Sigma + \frac{1}{2}\sum_{K \in \mathcal{T}_{h}}\int_{K} \delta^{n+\gamma} \nabla\mathbf{v}_{I}^{n+\gamma}\mathbf{n}^{n+\gamma} \cdot \mathbf{n}^{n+\gamma} w d\Omega \nonumber \\
&+& \frac{1}{2}\sum_{e \in \mathcal{E}} \int_{e} \left\{\left\{\mathbf{v}_{I}^{n+\gamma} \cdot \mathbf{n}^{n+\gamma}\right\}\right\} \left[\left[\mathbf{n}^{n+\gamma}\delta^{n+\gamma}w\right]\right]d\Sigma \nonumber \\
&-& \frac{1}{2}\sum_{e \in \mathcal{E}} \int_{e} \left\{\left\{\mathbf{n}^{n+\gamma}\delta^{n+\gamma}\right\}\right\} : \left[\left[\mathbf{v}_{I}^{n+\gamma} \otimes \mathbf{n}^{n+\gamma}w\right]\right] \nonumber \\
&=& \sum_{K \in \mathcal{T}_{h}}\int_{K} \frac{\delta^{n}}{\gamma \Delta t}w d\Omega - \frac{1}{2}\sum_{K \in \mathcal{T}_{h}}\int_{K} \mathbf{v}_{I}^{n} \cdot \nabla\delta^{n} w d\Omega \\
&-& \frac{1}{2}\sum_{e \in \mathcal{E}} \int_{e} \left\{\left\{\delta^{n}\mathbf{v}_{I}^{n}\right\}\right\} \cdot \left[\left[w\right]\right]d\Sigma - \frac{1}{2}\sum_{e \in \mathcal{E}} \int_{e} \left\{\left\{\mathbf{v}_{I}^{n}\right\}\right\} \cdot \left[\left[\delta^{n}w\right]\right]d\Sigma \nonumber \\
&-& \frac{1}{2}\sum_{e \in \mathcal{E}} \int_{e} \frac{\lambda^{n}}{2}\left[\left[\delta^{n}\right]\right] \cdot \left[\left[w\right]\right]d\Sigma - \frac{1}{2}\sum_{K \in \mathcal{T}_{h}}\int_{K} \delta^{n} \mathbf{n}^{n} \otimes \mathbf{n}^{n} : \nabla\mathbf{v}_{I}^{n} w d\Omega \nonumber \\
&-& \frac{1}{2}\sum_{e \in \mathcal{E}} \int_{e} \left\{\left\{\mathbf{v}_{I}^{n} \cdot \mathbf{n}^{n}\right\}\right\} \cdot \left[\left[\mathbf{n}^{n}\delta^{n}w\right]\right]d\Sigma \nonumber \\
&+& \frac{1}{2}\sum_{e \in \mathcal{E}} \int_{e} \left\{\left\{\mathbf{n}^{n}\delta^{n}\right\}\right\} : \left[\left[\mathbf{v}_{I}^{n}\otimes \mathbf{n}^{n}w\right]\right] \qquad \forall w \in X_{h}. \nonumber
\end{eqnarray}
The second TR-BDF2 stage can be described in a similar manner according to the formulations reported in \eqref{eq:second_stage_delta_wrong}, \eqref{eq:second_stage_delta}, and \eqref{eq:second_stage_delta_new}.

\subsection{Rising bubble benchmark}
\label{ssec:rising_bubble}

The two-dimensional rising bubble is a well-established benchmark for incompressible two-phase flows \cite{hysing:2009}. Moreover, this test case is particularly relevant since the motion of a single bubble is tracked and, therefore, the perimeter (that, in two dimensions, plays the role of the interfacial area) can be computed accurately. Two configurations are considered, whose physical relevant parameters and non-dimensional numbers are listed in Table \ref{tab:rising_bubble_param} and Table \ref{tab:rising_bubble_adim_param}, respectively. The initial configuration consists of a circular bubble of radius \(r_{0} = \SI{0.25}{\meter}\) centered at \(\left(x_{0},y_{0}\right) = \left(0.5, 0.5\right)\) in the rectangular domain \(\Omega = \left(0,1\right) \times \left(0,2\right)\). The density of the bubble is smaller than that of the surrounding fluid. The final time is \(T_{f} = \SI{3}{\second}\). No-slip boundary conditions are imposed on the top and bottom boundaries, whereas periodic conditions are prescribed in the horizontal direction. The initial velocity field is zero.
\begin{table}[H]
	\centering
	\begin{tabular}{c c c c c c c} 
		\hline
		Test case & \(\rho_{1} \left[\SI{}{\kilogram\per\meter\cubed}\right]\) & \(\rho_{2} \left[\SI{}{\kilogram\per\meter\cubed}\right]\) & \(\mu_{1} \left[\SI{}{\kilogram\per\meter\per\second}\right]\) & \(\mu_{2} \left[\SI{}{\kilogram\per\meter\per\second}\right]\) & \(g \left[\SI{}{\meter\per\second\squared}\right]\) & \(\sigma \left[\SI{}{\kilogram\per\second\squared}\right]\) \\
		\hline
		Config. 1 & \(1000\) & \(100\) & \(10\) & \(1\) & \(0.98\) & \(24.5\) \\ 
		\hline
		Config. 2 & \(1000\) & \(1\) & \(10\) & \(0.1\) & \(0.98\) & \(1.96\) \\  
		\hline
	\end{tabular}
	\caption{Physical parameters defining the configurations from rising bubble test case (data from \cite{hysing:2009}.}
	\label{tab:rising_bubble_param}
\end{table} 

\begin{figure}[H]
	\centering
	\includegraphics[width=0.9\textwidth]{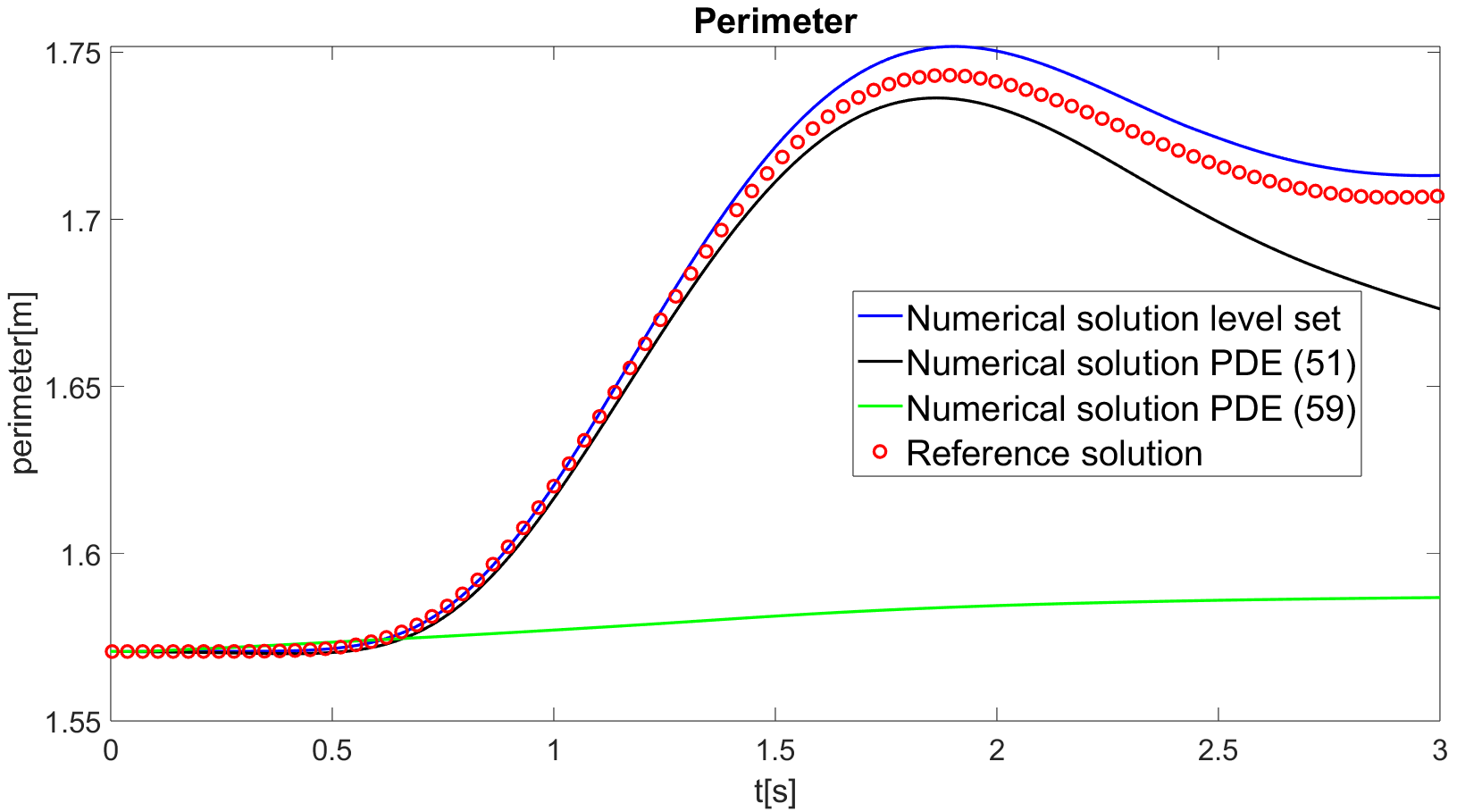}
	\caption{Rising bubble benchmark, configuration 1, evolution of the perimeter. The blue line denotes the perimeter computed from the level set function, the red dots report the perimeter computed from the reference solution in \cite{hysing:2009}, the black line represents the numerical solution obtained with \eqref{eq:delta_ev_7}, whereas the green line shows the numerical solution obtained with \eqref{eq:delta_ev_11}.}
	\label{fig:rising_bubble_case1}
\end{figure}

\begin{figure}[H]
	\centering
	\includegraphics[width=0.9\textwidth]{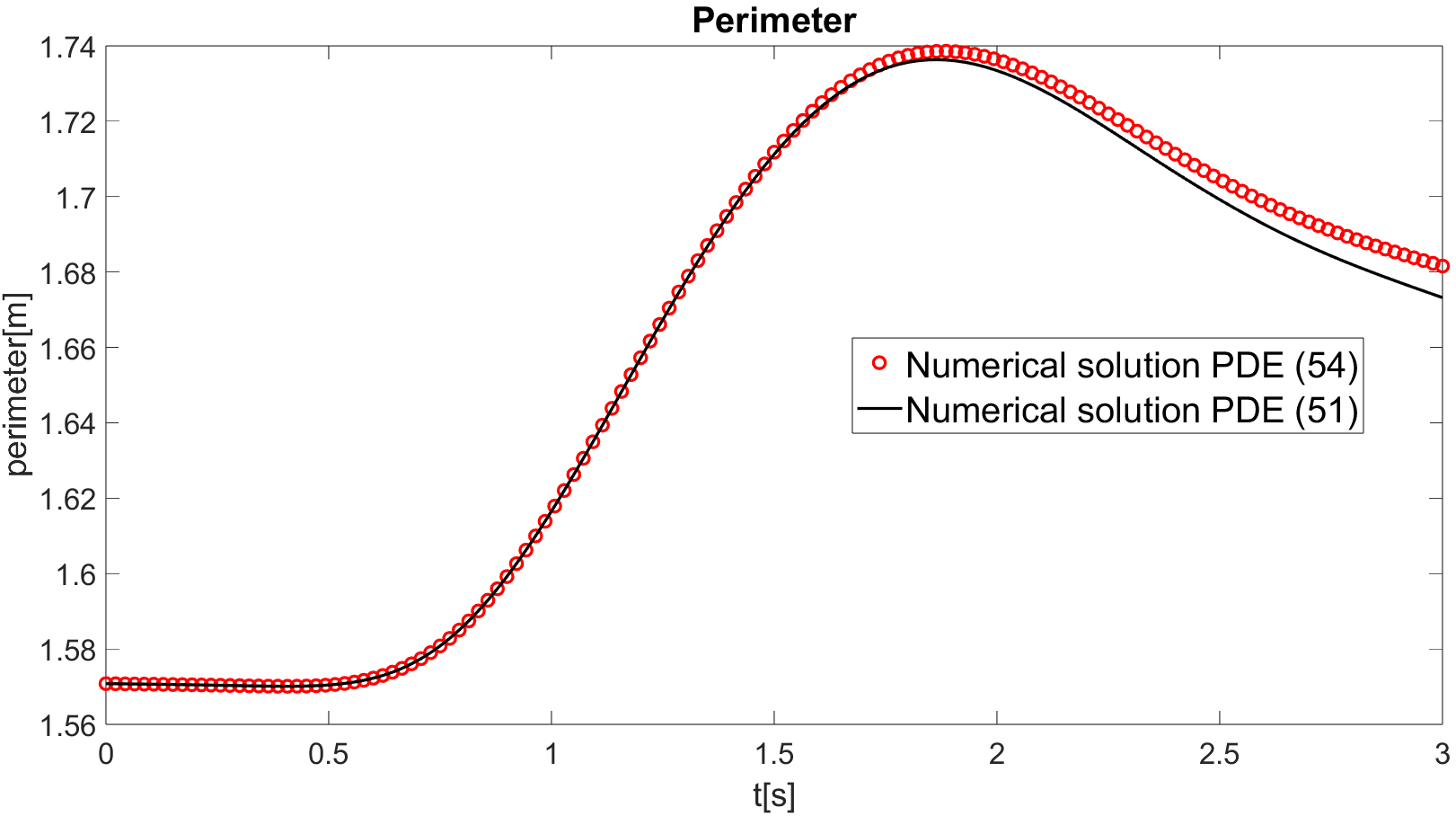}
	\caption{Rising bubble benchmark, configuration 1, evolution of the perimeter. The black line represents the numerical solution obtained with \eqref{eq:delta_ev_7}, whereas the red dots show the numerical solution obtained with \eqref{eq:delta_ev_9}.}
	\label{fig:rising_bubble_case1_perimeter_new}
\end{figure}

\begin{table}[H]
	\centering
	\begin{tabular}{c c c c c c} 
		\hline
		Test case & \(Re\) & \(Fr\) & \(We\) & \(\rho_{2}/\rho_{1}\) & \(\mu_{2}/\mu_{1}\) \\
		\hline
		Config. 1 & \(35\) & \(1\) & \(10\) & \(10^{-1}\) & \(10^{-1}\) \\ 
		\hline
		Config. 2 & \(35\) & \(1\) & \(125\) & \(10^{-3}\) & \(10^{-2}\) \\ 
		\hline
	\end{tabular}
	\caption{Non-dimensional numbers defining the configurations from rising bubble test case (data from \cite{hysing:2009}).}
	\label{tab:rising_bubble_adim_param}
\end{table}
We start with the first configuration. The computational grid is composed by \(320 \times 640\) elements, whereas the time step is \(\Delta t \approx 4.3 \cdot 10^{-3} \hspace{0.05cm} \SI{}{\second}\). Figure \ref{fig:rising_bubble_case1} shows the numerical results obtained using both formulation \eqref{eq:delta_ev_7} and \eqref{eq:delta_ev_11} in comparison with the perimeter computed from the level set function and the reference results in \cite{hysing:2009}. One can easily notice that \eqref{eq:delta_ev_11} leads to non reliable results, whereas a good qualitative agreement is established using \eqref{eq:delta_ev_7}. This further confirms the considerations reported in Section \ref{sec:interface_area}: while \eqref{eq:delta_ev_7} and \eqref{eq:delta_ev_11} are equivalent in the sense of distributions, they lead to significantly different numerical results and only a formulation which is valid for the whole space-time domain \(\Omega \times \left(0, T_{f}\right]\) should be used for numerical simulations. Figure \ref{fig:rising_bubble_case1_perimeter_new} shows instead a comparison between \eqref{eq:delta_ev_7} and \eqref{eq:delta_ev_9}. The computational results are very similar, showing that the two relations are equivalent in the whole space-time domain, as discussed in Section \ref{sec:interface_area}. 

\begin{figure}[H]
	\centering
	\includegraphics[width=0.9\textwidth]{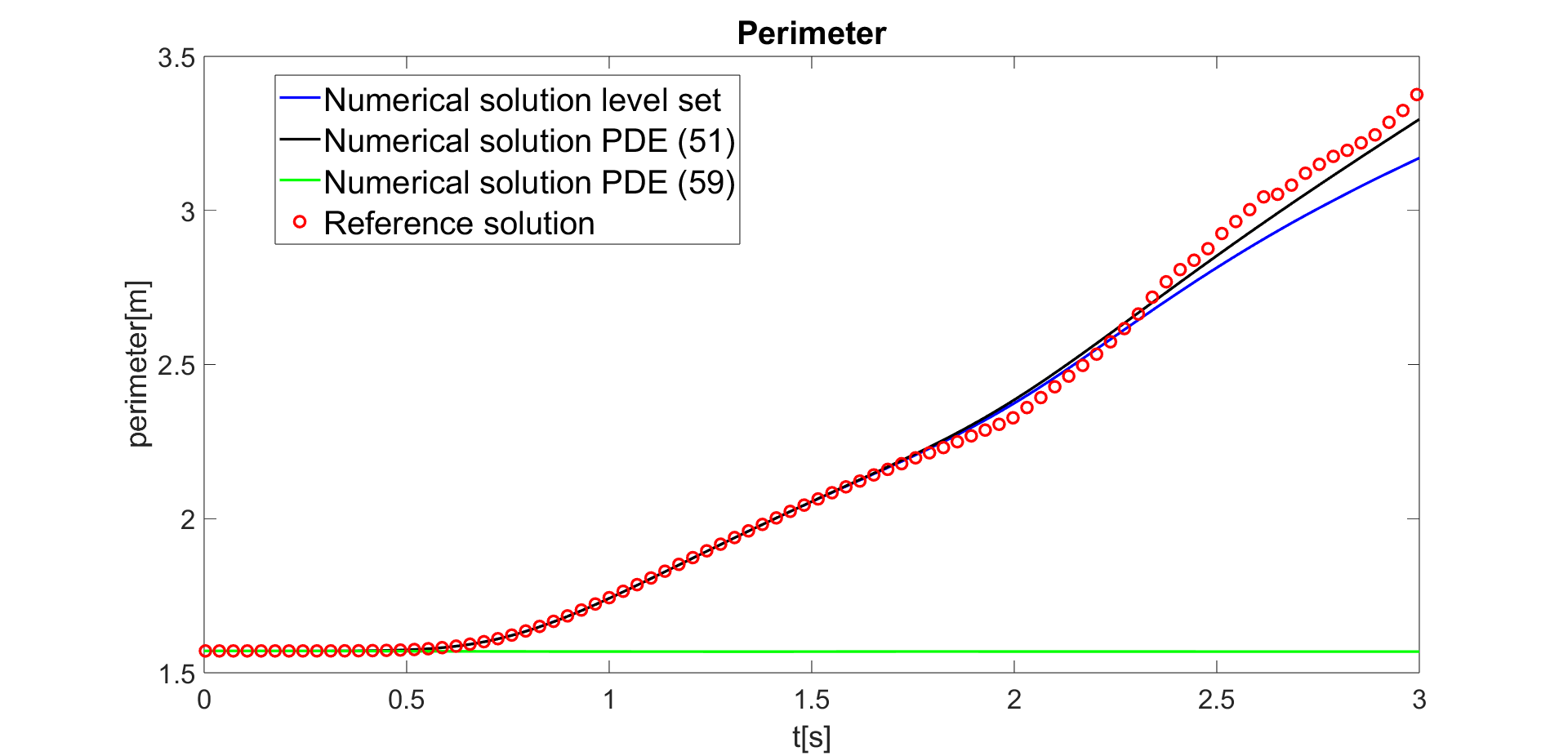}
	\caption{Rising bubble benchmark, configuration 2, evolution of the perimeter. The blue line denotes the perimeter computed from the level set function, the red dots report the perimeter computed from the reference solution in \cite{hysing:2009}, the black line represents the numerical solution obtained with \eqref{eq:delta_ev_7}, whereas the green line shows the numerical solution obtained with \eqref{eq:delta_ev_11}.}
	\label{fig:rising_bubble_case2}
\end{figure}

\begin{figure}[H]
	\centering
	\includegraphics[width=0.9\textwidth]{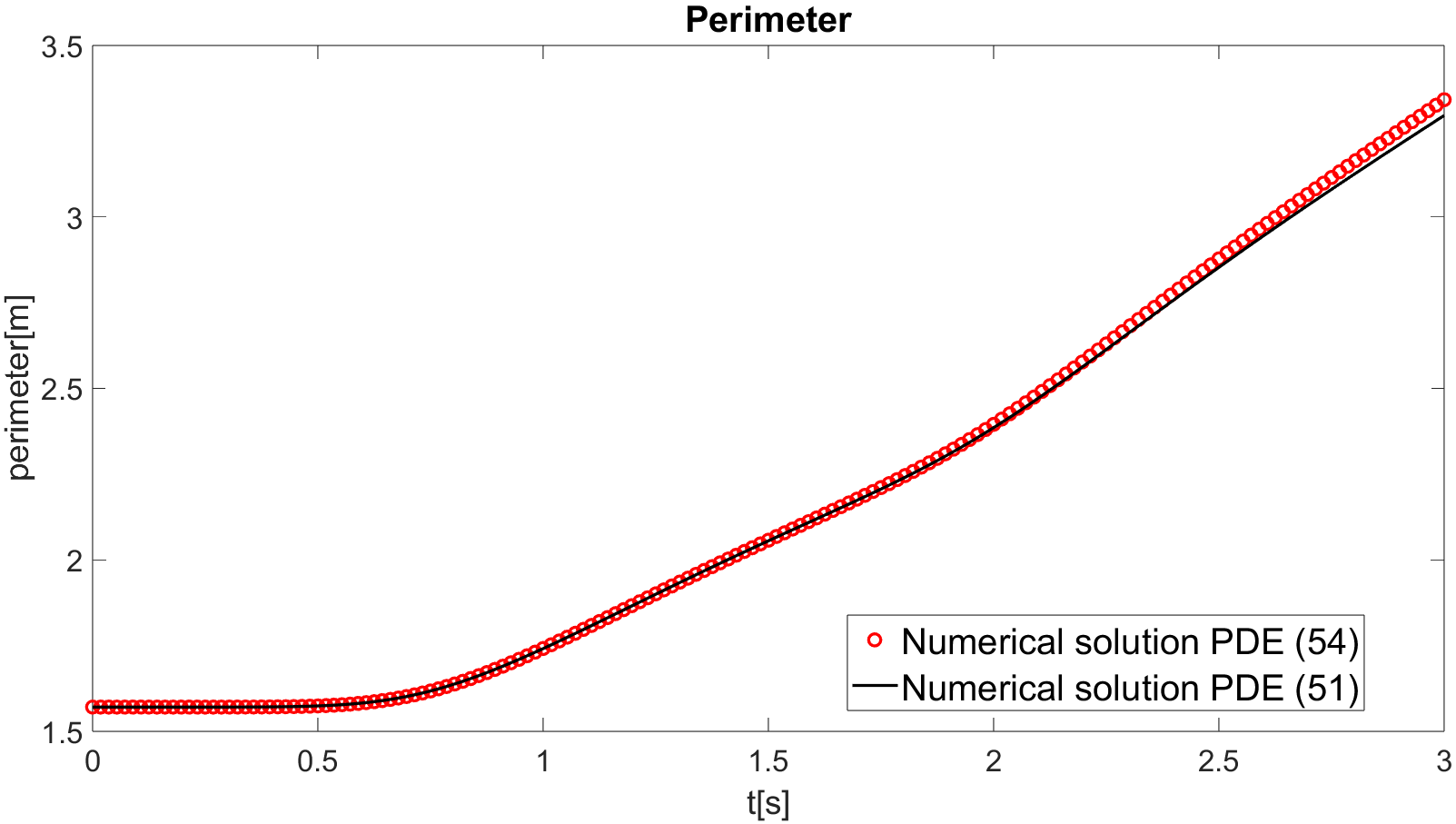}
	\caption{Rising bubble benchmark, configuration 2, evolution of the perimeter. The black line represents the numerical solution obtained with \eqref{eq:delta_ev_7}, whereas the red dots show the numerical solution obtained with \eqref{eq:delta_ev_9}.}
	\label{fig:rising_bubble_case2_perimeter_new}
\end{figure}

We analyze now the second configuration. This configuration is more challenging since the density of the bubble is much lower and it develops a non-convex shape with thin filament regions \cite{hysing:2009, orlando:2023b}. The time step is \(\Delta t \approx 3.6 \cdot 10^{-3} \hspace{0.05cm} \SI{}{\second}\). Figure \ref{fig:rising_bubble_case2} shows the numerical results obtained using both formulation \eqref{eq:delta_ev_7} and \eqref{eq:delta_ev_11} in comparison with the perimeter computed from the level set function and the reference results in \cite{hysing:2009}. The same considerations of the previous configuration hold and, in particular, one can easily notice an excellent agreement between the reference solution and the numerical results obtained using \eqref{eq:delta_ev_7}. Hence, this kind of relations are able to provide reliable values for the perimeter even when the interface undergoes large deformations and stretching, which are modelled by the terms on the right-hand side of \eqref{eq:delta_ev_7} and \eqref{eq:delta_ev_9}. Finally, Figure \ref{fig:rising_bubble_case2_perimeter_new} shows a comparison between \eqref{eq:delta_ev_7} and \eqref{eq:delta_ev_9}. One can easily notice once more that very similar results are obtained. We can empirically conclude from the numerical results that the right-hand side of \eqref{eq:delta_ev_7} and \eqref{eq:delta_ev_9} take into account stretching and deformation phenomena that increase the perimeter of the bubble.

\section{Conclusions}
\label{sec:conclu}

We have analyzed the evolution equations for a set of geometrical quantities that characterize the interface in two-phase flows. Several analytical relations have been presented, clarifying the hypotheses under which each equation is valid. More specifically, we have first reviewed the local balance laws which model two-phase flows and we have highlighted that interfacial source terms are proportional to the interfacial area density. Then, we have analyzed transport equations for the interfacial area density, for the unit normal vector and for the mean curvature. In particular, we have derived evolution equations in which the advecting field for these quantities is the complete interfacial velocity. This kind of relations can be important for applications where the interface is not well resolved. Finally, we have verified numerically the model equations on the classical rising bubble benchmark, showing their significantly different behaviour in a numerical framework. In future work, we plan to incorporate the relations for the geometrical quantities into Navier-Stokes equations and multifluid Baer-Nunziato type models for two-phase flows in order to compute terms related to surface tension and, more in general, to phase exchange.  

\section*{Acknowledgments}

The authors gratefully acknowledge N. Parolini for providing the original data of the rising bubble test case discussed in Section \ref{ssec:rising_bubble}. L.B. and G.O. have been partially supported by the ESCAPE-2 project, European Union’s Horizon 2020 Research and Innovation Programme (Grant Agreement No. 800897).

\appendix
\section{Equivalence between evolution equations of mean curvature}
\label{sec:mean_curvature_equations_equivalence}

In this Appendix, we prove the equivalence between \eqref{eq:mean_curvature_evolution_tris} and \eqref{eq:mean_curvature_evolution_bis}. We first notice that
\begin{equation}
\nabla_{s}\left(\mathbf{v}_{I} \cdot \mathbf{n}\right) = \nabla\left(\mathbf{v}_{I} \cdot \mathbf{n}\right) - \left[\nabla\left(\mathbf{v}_{I} \cdot \mathbf{n}\right) \cdot \mathbf{n}\right]\mathbf{n} = \left(\mathbf{I} - \mathbf{n} \otimes \mathbf{n}\right)\nabla\left(\mathbf{v}_{I} \cdot \mathbf{n}\right)	
\end{equation}
and that
\begin{eqnarray}
\dives\left[\nabla_{s}\left(\mathbf{v}_{I} \cdot \mathbf{n}\right)\right] &=& \left(\mathbf{I} - \mathbf{n} \otimes \mathbf{n}\right) : \nabla\left[\nabla_{s}\left(\mathbf{v}_{I} \cdot \mathbf{n}\right)\right] \nonumber \\
&=& \left(\mathbf{I} - \mathbf{n} \otimes \mathbf{n}\right) : \nabla\left[\left(\mathbf{I} - \mathbf{n} \otimes \mathbf{n}\right)\nabla\left(\mathbf{v}_{I} \cdot \mathbf{n}\right)\right]. 
\end{eqnarray}
Since
\begin{eqnarray}
\nabla\left[\left(\mathbf{I} - \mathbf{n} \otimes \mathbf{n}\right)\nabla\left(\mathbf{v}_{I} \cdot \mathbf{n}\right)\right] &=& -\nabla\mathbf{n}\left[\nabla\left(\mathbf{v}_{I} \cdot \mathbf{n}\right) \cdot \mathbf{n}\right] - \mathbf{n} \otimes \left(\nabla\mathbf{n}\right)^{T}\nabla\left(\mathbf{v}_{I} \cdot \mathbf{n}\right) \nonumber \\
&+& \left(\mathbf{I} - \mathbf{n} \otimes \mathbf{n}\right)\nabla\left[\nabla\left(\mathbf{v}_{I} \cdot \mathbf{n}\right)\right],
\end{eqnarray}
we obtain
\begin{eqnarray}
\left(\mathbf{I} - \mathbf{n} \otimes \mathbf{n}\right) : \nabla\left[\left(\mathbf{I} - \mathbf{n} \otimes \mathbf{n}\right)\nabla\left(\mathbf{v}_{I} \cdot \mathbf{n}\right)\right] &=& - 2H\nabla\left(\mathbf{v}_{I} \cdot \mathbf{n}\right) \cdot \mathbf{n} \\
&+& \left(\mathbf{I} - \mathbf{n} \otimes \mathbf{n}\right) : \nabla\left[\nabla\left(\mathbf{v}_{I} \cdot \mathbf{n}\right)\right]. \nonumber
\end{eqnarray}
If we substitute into \eqref{eq:mean_curvature_evolution_tris}, we obtain
\begin{eqnarray}\label{eq:mean_curvature_evolution_equaivalence_proof_1}
\frac{\partial_{s} H}{\partial t} &=& \left(2H^{2} - K\right)\left(\mathbf{v}_{I} \cdot \mathbf{n}\right) + H\nabla\left(\mathbf{v}_{I} \cdot \mathbf{n}\right) \cdot \mathbf{n} \\
&-& \frac{1}{2}\left(\mathbf{n} \otimes \mathbf{n} - \mathbf{I}\right) : \nabla\left[\nabla\left(\mathbf{v}_{I} \cdot \mathbf{n}\right)\right]. \nonumber
\end{eqnarray}
Comparing \eqref{eq:mean_curvature_evolution_equaivalence_proof_1} with \eqref{eq:mean_curvature_evolution_bis}, since \(\frac{\partial_{s}H}{\partial t} = \frac{\partial H}{\partial t} + \left(\mathbf{v}_{I} \cdot \mathbf{n}\right)\mathbf{n} \cdot \nabla H\), we notice that the equivalence between \eqref{eq:mean_curvature_evolution_tris} and \eqref{eq:mean_curvature_evolution_bis} is established if
\begin{equation}
\frac{1}{2}\nabla\mathbf{n} : \nabla\left[\left(\mathbf{v}_{I} \cdot \mathbf{n}\right)\mathbf{n}\right]^{T} - \frac{1}{2}\left(\nabla\mathbf{n}\right)\mathbf{n} \cdot \nabla\left(\mathbf{v}_{I} \cdot \mathbf{n}\right) = \left(2H^{2} - K\right)\left(\mathbf{v}_{I} \cdot \mathbf{n}\right).
\end{equation}
Starting from \eqref{eq:gaussian_curvature_def}, we notice that
\begin{equation}
K = 2H^{2} - \mathbf{n} \cdot \nabla H + \frac{1}{2}\left|\nabla \times \mathbf{n}\right|^{2} - \frac{1}{2}\mathbf{n} \cdot \left[\nabla\times\left(\nabla\times\mathbf{n}\right)\right]
\end{equation}
and, therefore, we get
\begin{eqnarray}
\left(2H^{2} - K\right)\left(\mathbf{v}_{I} \cdot \mathbf{n}\right) &=& \left(\mathbf{v}_{I} \cdot \mathbf{n}\right)\mathbf{n} \cdot \nabla H - \frac{1}{2}\left(\mathbf{v}_{I} \cdot \mathbf{n}\right)\left|\nabla \times \mathbf{n}\right|^{2} \nonumber \\
&+& \frac{1}{2}\left(\mathbf{v}_{I} \cdot \mathbf{n}\right)\mathbf{n} \cdot \left[\nabla \times \left(\nabla \times \mathbf{n}\right)\right]. 
\end{eqnarray}
Hence, \eqref{eq:mean_curvature_evolution_equaivalence_proof_1} reduces to
\begin{eqnarray}\label{eq:mean_curvature_evolution_equivalence_proof_2}
\frac{\partial H}{\partial t} &=& H\nabla\left(\mathbf{v}_{I} \cdot \mathbf{n}\right) \cdot \mathbf{n} - \frac{1}{2}\left(\mathbf{n} \otimes \mathbf{n} - \mathbf{I}\right) : \nabla\left[\nabla\left(\mathbf{v}_{I} \cdot \mathbf{n}\right)\right] \nonumber \\
&-& \frac{1}{2}\left(\mathbf{v}_{I} \cdot \mathbf{n}\right)\left|\nabla \times \mathbf{n}\right|^{2} + \frac{1}{2}\left(\mathbf{v}_{I} \cdot \mathbf{n}\right)\mathbf{n} \cdot \left[\nabla \times \left(\nabla \times \mathbf{n}\right)\right].
\end{eqnarray}
Since
\begin{equation}
\nabla \times \left(\nabla \times \mathbf{n}\right) = \nabla\left(\dive\mathbf{n}\right) - \dive\left(\nabla\mathbf{n}\right) = 2\nabla H - \dive\left(\nabla\mathbf{n}\right)
\end{equation}
and
\begin{equation}
\left|\nabla \times \mathbf{n}\right|^{2} = -\left[\dive\left(\nabla\mathbf{n}\right)\right] \cdot \mathbf{n} - \nabla\mathbf{n} : \left(\nabla\mathbf{n}\right)^{T},
\end{equation}
we obtain
\begin{eqnarray}\label{eq:mean_curvature_evolution_equivalence_proof_3}
\frac{\partial H}{\partial t} + \left(\mathbf{v}_{I} \cdot \mathbf{n}\right)\mathbf{n} \cdot \nabla H &=& H\nabla\left(\mathbf{v}_{I} \cdot \mathbf{n}\right) \cdot \mathbf{n} - \frac{1}{2}\left(\mathbf{n} \otimes \mathbf{n} - \mathbf{I}\right) : \nabla\left[\nabla\left(\mathbf{v}_{I} \cdot \mathbf{n}\right)\right] \nonumber \\
&+& \frac{1}{2}\left(\mathbf{v}_{I} \cdot \mathbf{n}\right)\nabla\mathbf{n} : \left(\nabla\mathbf{n}\right)^{T}.
\end{eqnarray}
Finally, since
\begin{equation}
\left(\mathbf{v}_{I} \cdot \mathbf{n}\right)\nabla\mathbf{n} : \left(\nabla\mathbf{n}\right)^{T} = \nabla\mathbf{n} : \nabla\left[\left(\mathbf{v}_{I} \cdot \mathbf{n}\right)\mathbf{n}\right]^{T} - \left(\nabla\mathbf{n}\right)\mathbf{n} \cdot \nabla\left(\mathbf{v}_{I} \cdot \mathbf{n}\right),
\end{equation}
we recover \eqref{eq:mean_curvature_evolution_bis}.

\pagebreak

\bibliographystyle{plain}
\bibliography{Interfacial_Quantities}

\end{document}